\documentclass[12pt, letterpaper]{extarticle}

\usepackage{helvet}

\usepackage{times}
\usepackage{comment}
\usepackage{placeins}
\usepackage{multirow}
\usepackage{rotating}

\usepackage{mathtools}
\usepackage{subfig}
\usepackage{arydshln}
\usepackage{longtable}
\usepackage{float}
\usepackage{multicol}
\usepackage{supertabular}
\usepackage{booktabs}
\usepackage[usenames,dvipsnames,svgnames,table]{xcolor}
\usepackage{longtable}

\usepackage[framemethod=TikZ]{mdframed}
\usepackage{fullpage}
\usepackage[switch]{lineno}

\usepackage{amsmath}
\usepackage{amssymb}
\usepackage{rotating}
\usepackage{array}
\usepackage{mathtools}
\usepackage[ruled]{algorithm2e}
\usepackage{algorithmic}
\usepackage{bm}
\usepackage{breqn}
\usepackage{comment}
\usepackage{enumitem}
\usepackage{graphics}
\usepackage{graphicx}
\usepackage{latexsym}
\usepackage{mathrsfs}
\usepackage{morefloats}
\usepackage{todonotes}
\usepackage{nicefrac}
\usepackage{authblk}

\title{The Preeminence of Ethnic Diversity in Scientific Collaboration}
\author[a,1]{Bedoor K AlShebli}
\author[a,b,1]{Talal Rahwan}
\author[a,c,1]{Wei Lee Woon}

\affil[a]{\normalsize 
Department of Computer Science, Masdar Institute, Khalifa University of Science and Technology, Abu Dhabi, P.O. Box 54224, UAE.}

\affil[b]{\normalsize 
Computer Science, New York University, Abu Dhabi, UAE.}

\affil[c]{\normalsize  
Expedia Inc., 333 108th AVE NE
Bellevue, WA 98004, USA.
\vspace*{0.5cm}}

\affil[1]{\footnotesize Joint corresponding authors. E-mail:\ \ bedoor@deeplearn.net;\ \ talal.rahwan@nyu.edu;\ \  wlwoon@deeplearn.net}
\date{}

\renewcommand{\bf}{}
\renewcommand{\textit}{}

\begin{document} 

\maketitle 

{\color{red}\textbf{IMPORTANT: The final version of this paper is published in Nature Communications 2018. We appreciate you not citing this arXiv version. }}

\begin{abstract}

{Inspired by the social and economic benefits of diversity, we analyze over 9 million papers and 6 million scientists to study the relationship between research impact and five classes of diversity: ethnicity, discipline, gender, affiliation, and academic age. Using randomized baseline models, we establish the presence of homophily in ethnicity, gender and affiliation. We then study the effect of diversity on scientific impact, as reflected in citations. Remarkably, of the classes considered, ethnic diversity had the strongest correlation with scientific impact. To further isolate the effects of ethnic diversity, we used randomized baseline models and again found a clear link between diversity and impact. To further support these findings, we use \textit{coarsened exact matching} to compare the scientific impact of ethnically diverse papers and scientists with closely-matched control groups. Here, we find that ethnic diversity resulted in an impact gain of 10.63\% for papers, and 47.67\% for scientists.}
\end{abstract}

\section*{Introduction}

Diversity is highly valued in modern societies \cite{wagner2017open, puritty2017without, page2008difference,
ager2013cultural,lee2014migrant, suedekum2014cultural}.
Social cohesion, tolerance and integration are linked to tangible benefits including economic vibrancy \cite{levine2014ethnic, herring2009does} and innovativeness \cite{lee2014migrant,paulus2016cultural,parrotta2014nexus,ostergaard2011does}. Far from being an abstract ideal, this conviction has guided many governmental and hiring policies and can have broad and long-lasting effects on society \cite{brown2015does,arcidiacono2015affirmative}. However, diversity is a complex issue, as groups can be diverse in terms of various attributes, such as ethnicity, gender, age and socioeconomic background. It is also unclear if all forms of diversity are beneficial. For instance, ethnic density has been associated with positive outcomes in terms of health \cite{alvarez2012health,das2010understanding}, while ethnic polarization has a negative effect on economic development \cite{montalvo2005economy}. Furthermore, diversity can be a divisive topic that is clouded by emotion, partisan loyalties and political correctness, all of which can hinder impartial discussions \cite{galinsky2015maximizing}. The factors above strongly motivate an objective study on the value of diversity, and on whether more diverse groups achieve greater success.

One domain in which this question can be effectively addressed is academia \cite{woolley2010evidence,hong2004groups}. The structure of academic collaboration is observable via co-authorships, which frequently involve scientists from different locations, disciplines and backgrounds \cite{jia2017quantifying, deville2014career}. Furthermore, academic output has an objective, widely-accepted measure---citation count \cite{sinatra2016quantifying,wang2013quantifying}.
This amenability to analysis has already attracted attempts at identifying the factors which underlie success in academia, an enterprise known as the Science of Science \cite{Fortunatoeaao0185}. 
Although many such factors have been studied, including gender \cite{nielsen2017opinion}, academic age \cite{jones2011age}, team size \cite{wuchty2007increasing}, interdisciplinarity \cite{uzzi2013atypical}, ethnicity \cite{freeman2015collaborating}, and affiliation \cite{jones2008multi,adams2013collaborations}, the study of these factors is extremely complex and many questions remain unanswered.

Our study seeks to address this shortcoming from a number of hitherto unexplored perspectives. Firstly, we compare homophily in scientific collaborations from the perspectives of age, gender, affiliation, and ethnicity.
We find clear signs of homophily in the cases of ethnicity, gender and affiliation. However, in only one case, ethnicity, was homophily was found to be increasing steadily over time.
Secondly, we examine the relationship between various classes of diversity and research impact at the level of scientific fields. Remarkably, we found that ethnic diversity is the most strongly associated with scientific impact. 
Thirdly, we compare the benefits of diversity on groups vs.~individuals, and find that the former outweighs the latter. Finally, we study the evolution and effect of diversity over time, team size, and number of collaborators, and verify that the above findings persist across all of these dimensions. The results of these multiple angles of analysis are combined to form a far richer picture of diversity than has been possible in the past.

\section*{Results}

\subsection*{Exploring Homophily}

A natural starting point for our study of diversity is to establish the extent to which \textit{homophily} \cite{mcpherson2001birds} exists in academia---i.e., whether scientists tend to collaborate more frequently with similar others---which would lead to an overall lack of diversity in scientific collaborations. We use the Microsoft Academic Graph dataset\footnote{https://www.microsoft.com/en-us/research/project/microsoft-academic-graph/}, and analyze 1,045,401 multi-authored papers (see Supplementary Figure 1 for the distribution of papers by year), written by 1,529,279 scientists, spanning 8 main fields and 24 subfields of science. We analyze diversity in terms of these five attributes: ethnicity ($\mathit{eth}$), discipline ($\mathit{dsp}$), gender ($\mathit{gen}$), affiliation ($\mathit{aff}$), and academic age ($\mathit{age}$); see Supplementary Note~1. Here, the abbreviations in parentheses are used in subsequent mathematical expressions to indicate the  associated attribute.
These attributes reflect many technical and social factors that influence teamwork and collaboration. \textit{Affiliation} indicates the geographic location, and may even reflect the way collaborative work is carried out---from the style and culture of collaboration to its mundane details, such as the medium used to collaborate, e.g., face-to-face interactions vs.~telecommunication or email. \textit{Academic age} is not only indicative of the amount of experience that a scientist has, but is also typically associated with actual age. \textit{Discipline} may reflect a scientist's substantive knowledge and his/her acquired skills through training, as well as the culture in which collaborative work is carried out. Finally, \textit{ethnicity} and \textit{gender} may play a role in shaping scientists' social identities, knowledge, and biases. To quantify diversity in terms of any of the aforementioned attributes, we use the \textit{Gini Impurity} \cite{bishop2006pattern}, resulting in the following group diversity indices, $d^G_{\mathit{eth}}$, $d^G_{\mathit{age}}$, $d^G_{\mathit{gen}}$, $d^G_{\mathit{dsp}}$ and $d^G_{\mathit{aff}}$ (an alternative diversity measure was also considered; see Supplementary Note~2 and Supplementary Figure~2).

To explore homophily, we generate different \textit{randomized baseline models} whereby a particular attribute---be it ethnicity, gender, affiliation or academic age---is shuffled.
For example, in the case of ethnicity, this process is akin to creating a universe in which ethnicity is disregarded in the selection of co-authors, while retaining other criteria.
To preserve the conditional distributions of the ethnicities, the shuffling process is constrained to only occur between authors of papers that have the same subfield, publication year, and number of authors; for full details, see Supplementary Note~3. 
This way, for every paper $p$ in the real dataset, there exists a matching paper $p'$ in the randomized dataset that may differ from $p$ in terms of ethnic diversity, but is identical to $p$ in terms of gender, affiliation, academic age, citations, publication year, and number of authors per paper. Importantly, while such a baseline model may produce homogeneous groups, the emergence of such groups is purely the result of random chance rather than homophily. As such, by comparing the real dataset with this baseline model, we can determine whether homophily exists, and if so, quantify the degree to which it is spread across academia. Figure~\ref{fig:HomophilyEffect:new}A compares our real dataset with the randomized baseline model in terms of the cumulative distributions of $d^G_x:x\in\{\mathit{eth},\mathit{age},\mathit{gen},\mathit{aff}\}$. As can be seen, for $x\in\{\mathit{eth},\mathit{gen},\mathit{aff}\}$, groups with low $d^G_x$ are more common in reality than would be expected by random chance, highlighting the fact that homophily does indeed exist in academia in terms of ethnicity, gender, and affiliation. However, for $x = \mathit{age}$, the opposite was observed (see Supplementary Figures~3-6 for subfield-specific distributions). These observations persist, regardless of the publication year (Figure~\ref{fig:HomophilyEffect:new}B), and the number of authors per paper (Figure~\ref{fig:HomophilyEffect:new}C). The temporal trends observed in Figure~\ref{fig:HomophilyEffect:new}B are particularly intriguing. For $d^G_{\mathit{eth}}$, while the population of scientists is becoming more ethnically diverse (see the steady increase in the red line), this trend is not reflected in the actual coauthor groupings, implying that ethnic homophily is steadily \textit{increasing}. For $d^G_{\mathit{age}}$, the actual level of diversity is greater than would be expected by random chance; this pattern is regularly observed in academia, e.g., consider the many publications resulting from advisor-advisee collaborations. For $d^G_{\mathit{gen}}$, although gender homophily continues to exist, it steadily \textit{decreases} over time, suggesting that women are playing an ever greater role in scientific endeavors. Finally, for $d^G_{\mathit{aff}}$, there is a marked \textit{decrease} in affiliation homophily around the 1990s; this is consistent with the jump in multi-university collaborations in the 1990s due to the widespread of the Internet and other technologies that facilitate collaboration across geographically distant scientists \cite{jones2008multi}.

\subsection*{The link between diversity and scientific impact}
Having explored homophily in academia, we now study the effects of homophily (and diversity) on research impact, measured by the number of citations received within five years of publication, denoted by $c^G_5$ (see Supplementary Note~4 and Supplementary Figure~7). Using the same dataset and notation described earlier, we study the relationship between a subfield's diversity and its academic impact. 
Here, we distinguish between two notions of diversity. The first is where the unit of analysis is a paper's \textit{set of authors}, while the second is where the unit of analysis is an individual scientist's \textit{entire set of collaborators}. We refer to the former as \textit{group diversity}, and to the latter as \textit{individual diversity}; see Figure~\ref{fig:indv_vs_grp} for an illustration comparing the two notions.

For each subfield, Figure~\ref{fig:groupindices_subfields}A  depicts the mean group diversity indices, $\big<d^G_x\big>:x\in\{\mathit{eth},\mathit{age},\mathit{gen},\mathit{dsp},\mathit{aff}\}$, against the mean five-year citation count, $\big<c_5^G\big>$, taken over papers in that subfield (notation summary and formal definitions are in Supplementary Table~1 and Supplementary Note~2, respectively). Remarkably, we find that a subfield's ethnic diversity is the most strongly correlated with impact ($r=0.77$); the positive correlation persists even when the subfields are studied in isolation (Supplementary Figures~8 and Supplementary Table~2), regardless of the number of authors per paper (Supplementary Figure~9). 
These findings are further supported by the regression analysis in Table~\ref{tab:RegressionCEMFactors_MainField}. 
While these findings do not imply causation, it is still suggestive that one can largely predict scientific impact based solely on average ethnic diversity, especially given that ethnicity is arguably unrelated to technical competence.

Having studied group diversity, we now move our attention to individual diversity. Here, we analyze scientists with at least 10 collaborators each, amounting to a total of 5,103,877 collaborators over 9,472,439 papers (see Supplementary Table~3 for a summary of all filters applied on the dataset). For each subfield, Figure~\ref{fig:groupindices_subfields}B depicts the mean individual diversity indices, $\big<d^I_x\big>:x\in\{\mathit{eth},\mathit{age},\mathit{gen},\mathit{dsp},\mathit{aff}\}$, against the mean five-year citation count, $\big<c_5^I\big>$, taken over \textit{scientists} in that subfield. As can be seen, a subfield's ethnic diversity is again the most strongly correlated with impact ($r=0.55$), even when the subfields are studied in isolation (Supplementary Figure~10 and Supplementary Table~4).

The above results highlight a potential dysfunction. While homophily was observed for ethnicity, affiliation and gender, the only attribute for which it was found to be \textit{increasing} over time was ethnicity, which seems strange given the apparent preeminence of ethnic diversity. Motivated by this observation, we further explore the relationship between ethnic diversity and scientific impact in the randomized universe used earlier in Figure~\ref{fig:HomophilyEffect:new}. Recall that, in such a universe, ethnicity is excluded as a criterion for selecting co-authors while the other factors are preserved. Hence, it stands to reason that any differences in impact between the randomized and real datasets can be attributed to ethnic diversity. To examine these differences, we partitioned the papers into two categories, labeled as diverse ($d^G_{\mathit{eth}}>\widetilde{d}^G_{\mathit{eth}}$) and non-diverse ($d^G_{\mathit{eth}}\leq \widetilde{d}^G_{\mathit{eth}}$) where the tilde denotes the median. The scientists were similarly partitioned into diverse ($d^I_{\mathit{eth}}>\widetilde{d}^I_{\mathit{eth}}$) and non-diverse ($d^I_{\mathit{eth}}\leq \widetilde{d}^I_{\mathit{eth}}$). We find that the diverse consistently outperforms the non-diverse, regardless of the year of publication (Figure~\ref{fig:PubYearandNumAuth}E), the number of authors per paper (Figure~\ref{fig:PubYearandNumAuth}G), and the number of collaborators per scientist (Figures~\ref{fig:PubYearandNumAuth}I). We replicated these plots using the randomized, instead of the real, dataset (Figures~\ref{fig:PubYearandNumAuth}F, \ref{fig:PubYearandNumAuth}H and \ref{fig:PubYearandNumAuth}J). As can be seen, the performance gap between the diverse and non-diverse almost entirely disappears in the randomized dataset, suggesting that the observed impact gains in the real dataset could indeed be attributed to ethnic diversity. Note that, in the real dataset, a large proportion of papers have $d^G_{\mathit{eth}} = 0$ (see Figure~\ref{fig:PubYearandNumAuth}A), and a large proportion of scientists have $d^I_{\mathit{eth}} = 0$ (see Figure~\ref{fig:PubYearandNumAuth}C). As such, the observed performance gap between the diverse and the non-diverse could be predominantly due to these papers and scientists being less impactful than their counterparts whose $d^G_{\mathit{eth}} > 0$ and $d^I_{\mathit{eth}} > 0$, respectively. To determine whether this is the case, we replicated the analysis of papers but after excluding those with $d^G_{\mathit{eth}} = 0$, and likewise replicated the analysis of scientists but after excluding those with $d^I_{\mathit{eth}} = 0$; see Supplementary Figure~11. As can be seen, even after this exclusion, the diverse mostly outperform the non-diverse, regardless of publication year, number of authors per paper, and number of collaborators per scientist.

\subsection*{Inferring causality}
To provide further evidence of the
link between ethnic diversity and scientific impact, we use \textit{coarsened exact matching}~\cite{iacus2012causal}, a technique typically used to infer causality in observational studies \cite{catalini2015incidence}. Specifically, it matches the control and treatment populations with respect to the confounding factors identified, thereby eliminating the effect of these factors on the phenomena under investigation. In our case, when studying \textit{group} ethnic diversity, the \textit{treatment set} consists of papers for which $d^G_{\mathit{eth}} > P_{100-i}\left(d^G_{\mathit{eth}}\right)$, and the \textit{control set} of papers for which $d^G_{\mathit{eth}} \leq P_i\left(d^G_{\mathit{eth}}\right)$, where $P_i\left(d^G_{\mathit{eth}}\right)$ denotes the $i^{th}$ percentile of $d^G_{\mathit{eth}}$. This process is repeated using $i=10, 20, 30, 40, 50$, corresponding to progressively larger gaps in ethnic diversity between the two populations. Thus, if ethnic diversity is indeed associated with increased scientific impact, we would expect to find a significant difference in impact between the two populations, and expect this difference to increase in tandem with the aforementioned gap in diversity. The confounding factors identified were the year of publication,  number of authors, field of study, authors' impact prior to publication, and university ranking. The same process was carried out for \textit{individual} ethnic diversity, for which the confounding factors were academic age, number of collaborators, discipline, and university ranking; see Supplementary Note~5 and Supplementary Figures~12 and 13 for more details, and Supplementary Figure~14 for an illustration of how this process works on a given collection of papers.
The results for group and individual ethnic diversities are summarized in Tables~\ref{tab:CEMGroup} and \ref{tab:CEMIndividual}, respectively. As can be seen, increasing the diversity gap between the control and treatment populations is often accompanied by a greater difference in scientific impacts between the two populations.  
Clearly, these results do not suggest that diversity is the only causal factor. For example, one may argue that highly-ranked universities tend to attract students from around the world and are more ethnically diverse as a result; indeed we verified that this was the case (see Supplementary Note~6 and Supplementary Figures~15 and 16). In such situations, coarsened exact matching is particularly useful precisely because it allows us to establish causality despite such effects.

\subsection*{Interplay between group and individual ethnic diversity}
Finally, we investigate the interplay between group ethnic diversity, $d^G_{\mathit{eth}}$, and individual ethnic diversity, $d^I_{\mathit{eth}}$. To this end, for each of the 1,045,401 papers in our dataset, we calculate $d^I_{\mathit{eth}}$ averaged over the authors in that paper; we denote this as $\big<d^I_{\mathit{eth}}\big>_{\textnormal{paper}}$. This allows us to study the ways in which the two notions of diversity vary in the same paper. Indeed, as illustrated in Figure~\ref{fig:DiversityIndicesExplained}, a paper can have high $d^G_{\mathit{eth}}$ and at the same time have low $\big<d^I_{\mathit{eth}}\big>_{\textnormal{paper}}$, and vice versa. With this in mind, we studied the impact, $\langle c^G_5\rangle$, of papers falling in different ranges of $d^G_{\mathit{eth}}$ and $\langle d^I_{\mathit{eth}}\rangle_{\textnormal{paper}}$; see the matrix at the bottom-right corner of Figure~\ref{fig:DiversityIndicesExplained}. Here, if we denote this matrix by $A$, and label the bottom row and leftmost column as 1, we find that $\sum_{i=1}^4 A_{i,1}< \sum_{i=1}^4 A_{1,i}$ and $\sum_{i=1}^4 A_{i,4}> \sum_{i=1}^4 A_{4,i}$. Hence, while it appears that both group and individual diversities can be valuable, the former seems to have a greater effect on scientific impact. In other words, having co-authors who are inclined to collaborate across ethnic lines (i.e., co-authors whose \textit{individual} ethnic diversity is high) appears to be not as important as the mere presence of co-authors of different ethnicities (i.e., co-authors whose \textit{group} ethnic diversity is high).

\section*{Discussion}
To summarize, this study is the first to cover five different classes of diversity, which allowed us to illuminate many interesting connections between diversity and scientific collaboration. It was also important to establish the occurrence of homophily, and 
this was achieved via a set of randomized baseline models. These were used to compare observed collaborations with simulated data where the attribute of interest was randomized while controlling for the relevant confounding variables. These comparisons revealed clear and consistent patterns of homophily in the cases of ethnicity, gender and affiliation, and also revealed that ethnicity was the only attribute for which homophily is increasing over time. In the case of academic age, inverse homophily was found, i.e., scientists seem to prefer collaborating with individuals from different age groups,
a possible reflection of the widely-held practice of research students being mentored by, and collaborating with, more senior academics.

Armed with these results, we shifted our focus to the \textit{effect} of homophily (and diversity) on scientific impact. This analysis was conducted using a number of different analytical tools, including regression analysis, randomized baseline models, and coarsened exact matching. Broadly, we found that diversity was positively correlated with impact, though the statistical significance of the observed effect varied significantly depending on the class of diversity and field of study. Overall, discipline and affiliation diversity were the least correlated with impact, a surprising finding given the apparent importance of these attributes.
Conversely, ethnic diversity had the strongest correlation, which is especially surprising since ethnicity is not as related to technical competence as the other classes mentioned.

These findings have significant implications. For one, recruiters should always strive to encourage and promote ethnic diversity, be it by recruiting candidates who complement the ethnic composition of existing members, or by recruiting candidates with proven track records in collaborating with people of diverse ethnic backgrounds. Another implication is that, while collaborators with different skill sets are often required to perform complex tasks, multidisciplinarity should not be an end in of itself; bringing together individuals of different ethnicities---with the attendant differences in culture and social perspectives---could ultimately produce a large payoff in terms of performance and impact. To put it differently, intangible factors such as team cohesion and a sense of \textit{esprit de corps} should be considered in addition to technical alignment.

The underlying message is an inclusive and uplifting one. In an era of increasing polarization and identity politics, our findings may positively contribute to the societal conversation and reinforce the conviction that good things happen when people of different backgrounds, cultures, and ethnicities come together to work towards shared goals and the common good.

\section*{Data availability}
The details of all data and methods used are given in Supplementary Note~1.


\section*{Acknowledgements}
We thank Steven Skiena and his team for providing access to their Name Ethnicity Classifier tool \cite{ambekar2009name, ye2017nationality}. We also thank Kinga Makovi for suggesting the use of coarsened exact matching for causal inference.

\section*{Author Contributions}
B.K.A., T.R. and W.L.W. conceived of and designed the experiments.
B.K.A. and W.L.W. performed the coding of the experiments.
B.K.A., T.R. and W.L.W. wrote the manuscript.
B.K.A. and T.R. produced the figures and tables.


\section*{Competing Interests} 
The authors declare no competing interests.
\clearpage

\renewcommand{\textbf}{}


\begin{figure*}
\centering
\includegraphics[width=\textwidth]{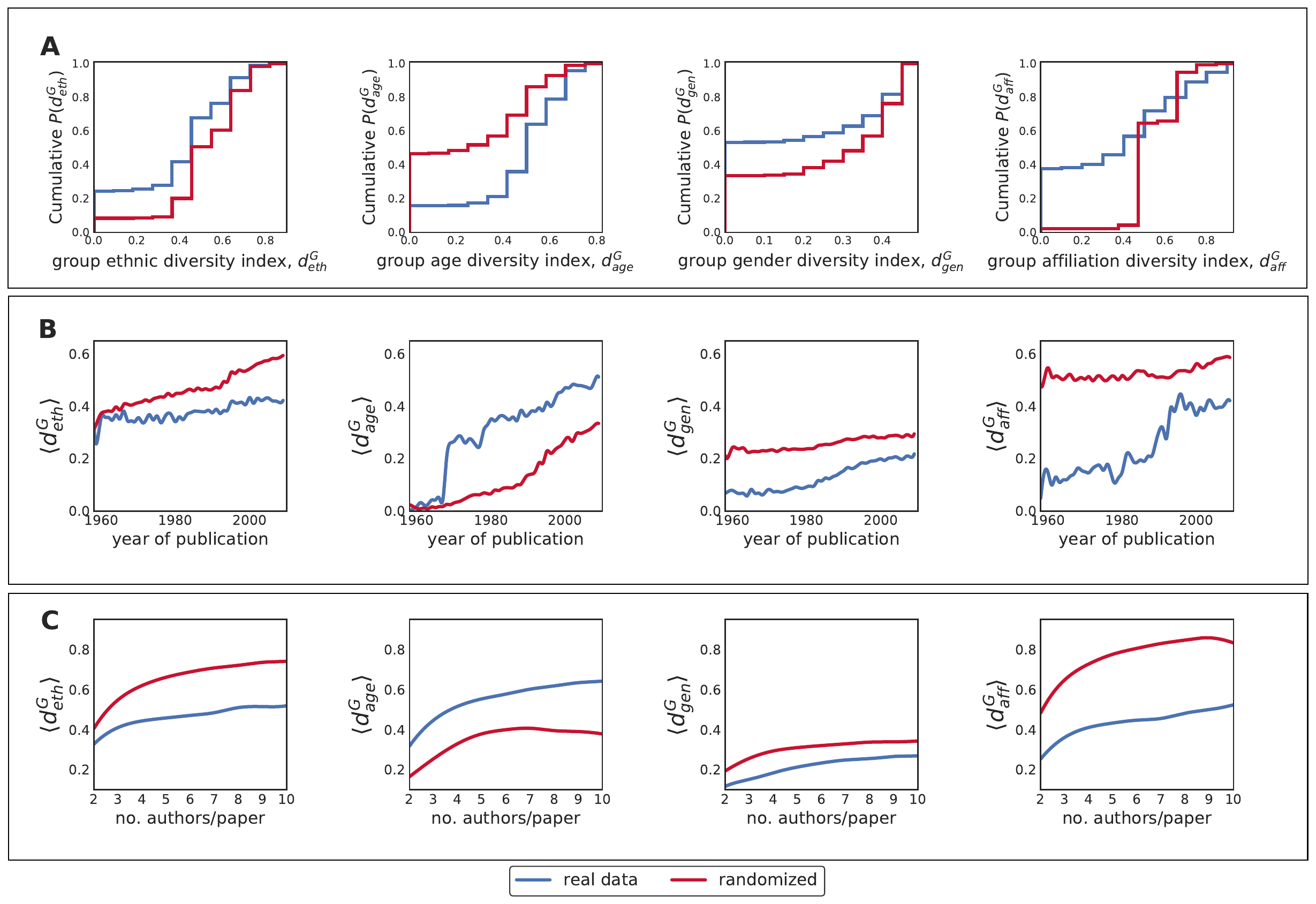}
\caption{\footnotesize
Exploring homophily in real vs. randomized data. Each column corresponds to a different class of diversity, and each row presents the results of a specific set of experiments whereby $d^G_x:x \in \{\mathit{eth}, \mathit{age}, \mathit{gen}, \mathit{aff}\}$ in real data is compared against randomized data. \textbf{(A)} Cumulative distributions of $d^G_x$. \textbf{(B)} Change in mean diversity $\langle d^G_x\rangle$ over time. \textbf{(C)} Mean diversity $\langle d^G_x\rangle$ for papers with different number of authors.
}
\label{fig:HomophilyEffect:new}
\end{figure*}

\begin{figure*}
\centering
\includegraphics[width=0.4\textwidth]{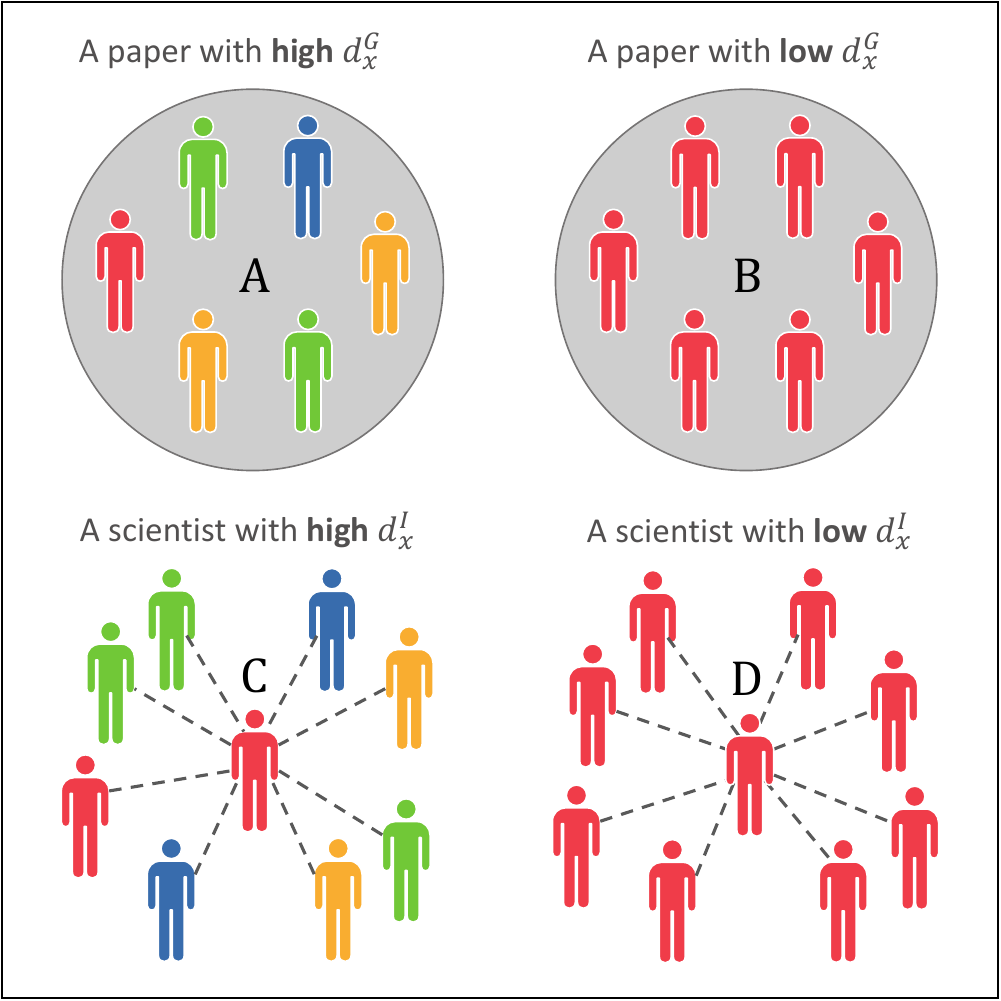}
\caption{\footnotesize Group vs.~individual diversity. For any given class of diversity, $x\in\{\mathit{eth},\mathit{age},\mathit{gen},\mathit{dsp},\mathit{aff}\}$, differences in color represent differences in terms of $x$. The group diversity index $d^G_x$ of Paper A is higher than that of Paper B. The individual diversity index of Scientist C is higher than that of Scientist D.}
\label{fig:indv_vs_grp}
\end{figure*}

\begin{figure*}
\centering
\includegraphics[width=1\textwidth]{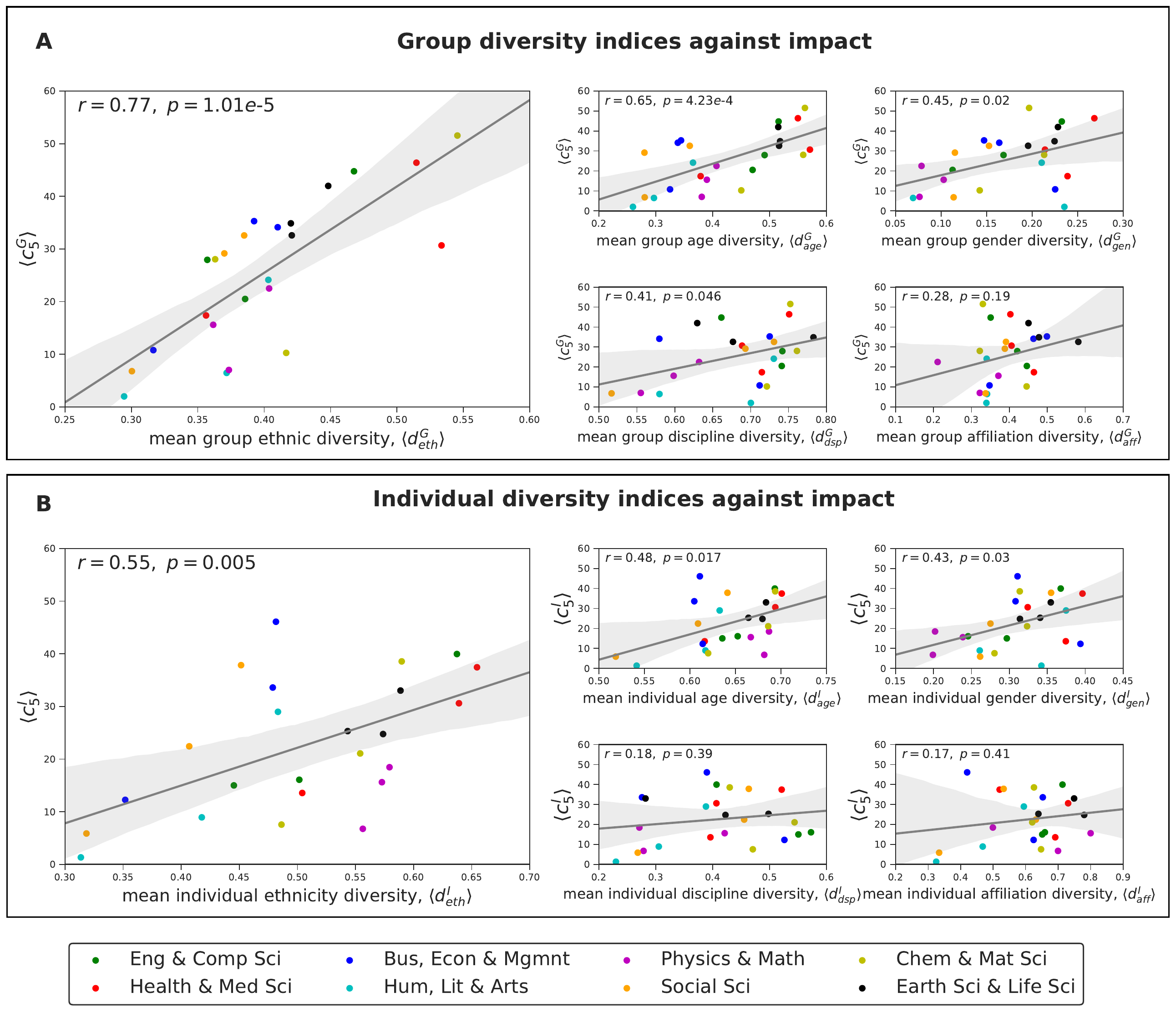}
\caption{\footnotesize 
Group and individual diversity vs.~impact in each subfield.
In each subplot, the points correspond to subfields, the color indicates the main field, while the solid line and the shaded area represent the regression line and the 95\% confidence interval, respectively. Each regression has also been annotated with the corresponding Pearson's $r$ and $p$ values. {\bf(A)} For each subfield, the subplots depict the mean \textit{group} diversity indices, $\langle d^G_{\mathit{eth}}\rangle$, $\langle d^G_{\mathit{age}}\rangle$, $\langle d^G_{\mathit{gen}}\rangle$, $\langle d^G_{\mathit{dsp}}\rangle$ and $\langle d^G_{\mathit{aff}}\rangle$, against the mean five-year citation count, $\langle c_5^G\rangle$, taken over \textit{papers} in that subfield. {\bf(B)} For each subfield, the subplots depict the mean \textit{individual} diversity indices, $\langle d^I_{\mathit{eth}} \rangle$, $\langle d^I_{\mathit{age}}\rangle $, $\langle d^I_{\mathit{gen}}\rangle $, $\langle d^I_{\mathit{dsp}}\rangle $ and $\langle d^I_{\mathit{aff}}\rangle $, against the mean five-year citation count, $\langle c_5^I\rangle $, taken over \textit{scientists} in that subfield. 
}
\label{fig:groupindices_subfields}
\end{figure*}

\begin{figure*}[!htb]%
\centering
\includegraphics[width=0.87\textwidth]{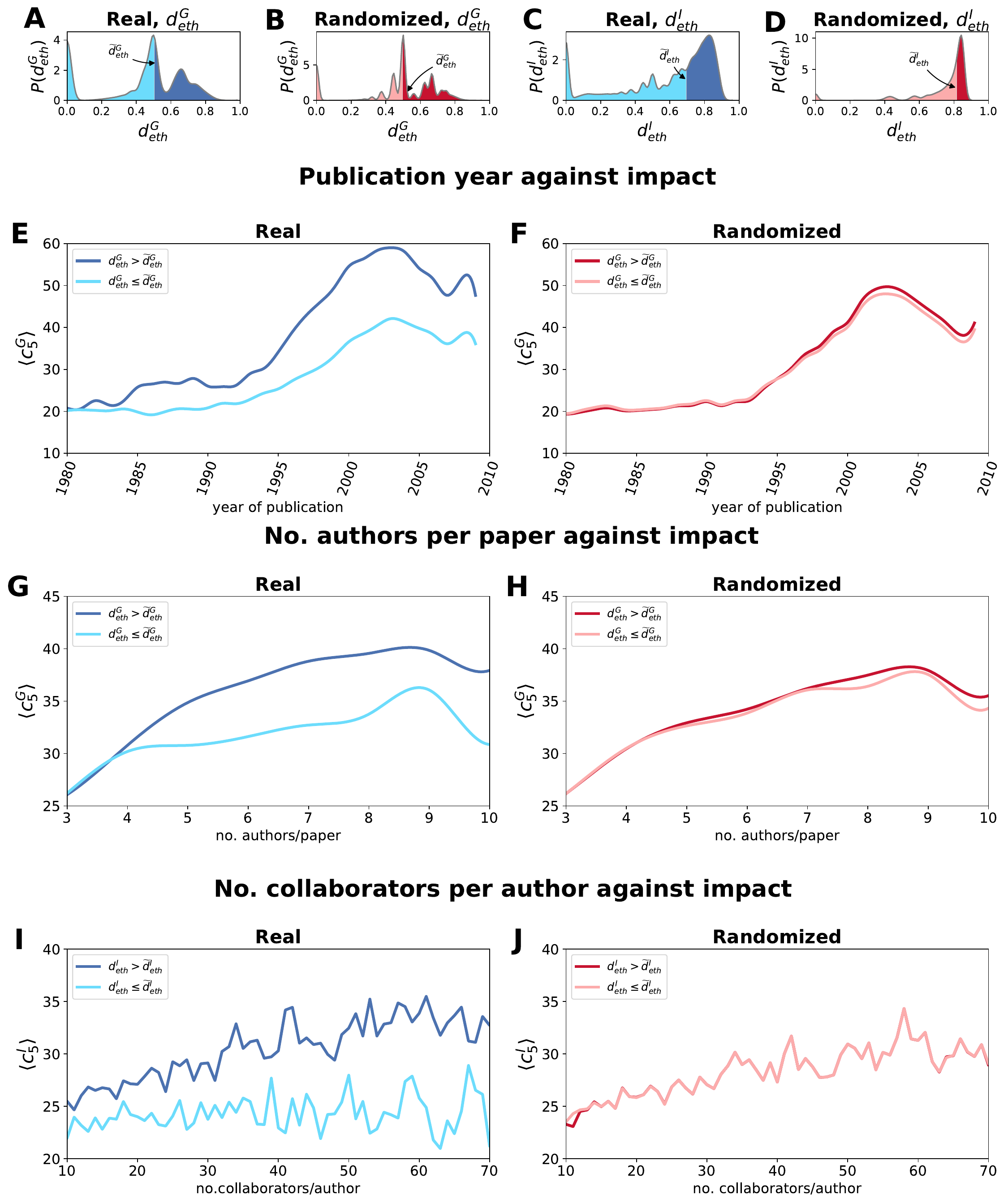}
\caption{\footnotesize The relationship between ethnic diversity and impact.
{\bf(A)} Distribution of $d^G_{\mathit{eth}}$ in real data. Papers were partitioned into two categories: \textit{diverse} (highlighted in the darker tones, with $d^G_{\mathit{eth}}>\widetilde{d}^G_{\mathit{eth}}$) and \textit{non-diverse}
(highlighted in the lighter tones, with $d^G_{\mathit{eth}}\leq\widetilde{d}^G_{\mathit{eth}}$), where the tilde denotes the median.
{\bf(B)} The same as Figure~\ref{fig:PubYearandNumAuth}A, but for randomized data. 
{\bf(C)} and {\bf(D)} The same as Figures~\ref{fig:PubYearandNumAuth}A and \ref{fig:PubYearandNumAuth}B, respectively, but with $d^I_{\mathit{eth}}$ instead of $d^G_{\mathit{eth}}$.  
{\bf(E)} $\langle c^G_5 \rangle$ against publication year in real data. 
{\bf(F)} The same as Figure~\ref{fig:PubYearandNumAuth}E, but for randomized data. 
{\bf(G)} $\langle c^G_5 \rangle$ against number of authors per paper in real data. 
{\bf(H)} The same as Figure~\ref{fig:PubYearandNumAuth}G, but for randomized data. 
{\bf(I)} $\langle c^I_5 \rangle$ against number of collaborators per scientist in real data. 
{\bf(J)} The same as Figure~\ref{fig:PubYearandNumAuth}I, but for randomized data.
}
\label{fig:PubYearandNumAuth}
\end{figure*}

\clearpage

\begin{figure*}[!htb]%
\centering
\includegraphics[width=13cm]{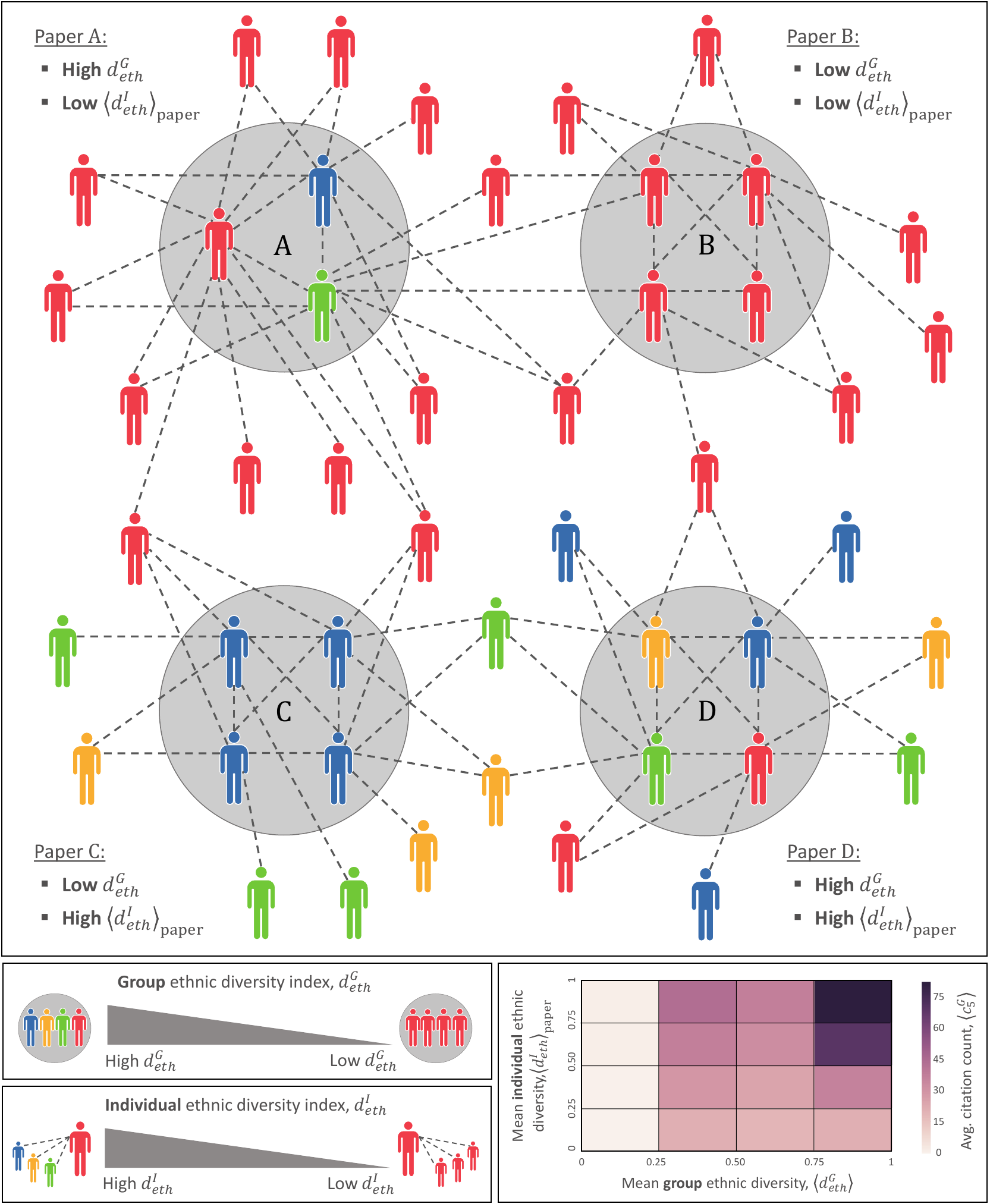}
\caption{\footnotesize The interplay between group and individual ethnic diversity.
The top part of the figure illustrates an example of 4 papers. The authors of paper~A have different ethnicities, but each has ethnically-homogeneous collaborators. Then, one could argue that paper~A has high $d^G_{\mathit{eth}}$ but low $\langle d^I_{\mathit{eth}}\rangle_{\textnormal{paper}}$. Similarly, paper~B has low $d^G_{\mathit{eth}}$ and low $\langle d^I_{\mathit{eth}}\rangle_{\textnormal{paper}}$, paper~C has low $d^G_{\mathit{eth}}$ and high $\langle d^I_{\mathit{eth}}\rangle_{\textnormal{paper}}$, and paper~D has high $d^G_{\mathit{eth}}$ and high $\langle d^I_{\mathit{eth}}\rangle_{\textnormal{paper}}$. The matrix at the bottom-right corner represents the mean citation counts, $\langle c^G_5\rangle$, of papers falling in different ranges of $d^G_{\mathit{eth}}$ and $\langle d^I_{\mathit{eth}}\rangle_{\textnormal{paper}}$.}
\label{fig:DiversityIndicesExplained}
\end{figure*}

\FloatBarrier
\begin{table}[H]
\caption{Regression analyses of diversity classes on academic impact.
The regression tables below present the effect of each of the five group diversity indices, $d^G_{\mathit{x}}: x\in\{\mathit{eth},\mathit{age},\mathit{gen},\mathit{aff},\mathit{dsp}\}$, on the paper's impact, $c^G_5$.
Along with each class of diversity, the following predictor variables were used: university ranking, author's prior impact, year of publication, and number of authors. Here, university rankings are based on the 2017 Academic Ranking of World Universities, also known as the Shanghai Ranking, whereas an author's prior impact is measured as the annual number of citations that he/she accumulated prior to the year in which the paper was published. The columns correspond to papers from different fields. Of the five classes of diversity studied, ethnic diversity (Table~(A)) was the only one for which all coefficients in the first row ($d^G_{\mathit{eth}}$) are positive and significant.
}
\label{tab:RegressionCEMFactors_MainField}
{
\begin{center}
\resizebox{\textwidth}{!}{%
\begin{tabular}{lcccccccc}
\multicolumn{9}{l}{\fontsize{13}{13}\selectfont{\textbf{\ \ (A)} Group ethnic diversity}}\\
\hline\smallskip
&\multicolumn{8}{c}{Citation Count, $c^G_5$}
\\\cline{2-9}
                             &    Engineering \&  &  Health \& Medical    &    Business, Economics     &    Humanities, Literature     &    Physics \&     &    Social     &    Chemical \&     &    Life Sciences \&      \\
                             &    Computer Science     & Sciences    &   \& Management     &    \& Arts     &    Mathematics     &    Sciences     &    Material Sciences     &    Earth Sciences      \\                             
\midrule
\midrule

$d^G_{\mathit{eth}}$                & 7.40***    & 3.00***   & 5.21***    & 4.77***    & 8.04**     & 4.39**     & 4.29**     & 3.94***     \\
        & (2.44)     & (0.64)    & (1.64)     & (1.79)     & (3.30)     & (1.89)     & (1.95)     & (1.45)      \\
University Ranking                 & -1.22***   & -1.08***  & -0.60**    & -0.52**    & -0.16      & -0.55*     & -0.35      & -1.35***    \\
                             & (0.39)     & (0.08)    & (0.24)     & (0.26)     & (0.46)     & (0.29)     & (0.29)     & (0.23)      \\
Author's Prior Impact & 0.62***    & 1.24***   & 1.52***    & 1.61***    & 0.72***    & 1.51***    & 1.60***    & 1.53***     \\
                             & (0.01)     & (0.01)    & (0.01)     & (0.01)     & (0.02)     & (0.01)     & (0.01)     & (0.01)      \\
Year of Publication         & 0.20     & 0.24***  & 0.07    & 0.48***    & 0.13     & 0.37**    & 0.24     & 0.24***   \\
                             & (0.21)   & (0.01)    & (0.10)   & (0.10)     & (0.16)   & (0.17)    & (0.17)   & (0.01)     \\
Number of Authors & 0.00       & 0.59***   & 0.23       & 0.27       & 1.06       & 0.46**     & 0.51***    & 0.69***     \\
                             & (0.27)     & (0.15)    & (0.17)     & (0.18)     & (1.03)     & (0.19)     & (0.19)     & (0.11)      \\
const                    & 2221.02*** & 598.55*** & 1081.71*** & 1085.84*** & 1289.91*** & 2142.17*** & 1813.42*** & 2750.75***  \\
                             & (270.36)   & (22.94)   & (114.27)   & (124.13)   & (230.16)   & (194.14)   & (188.89)   & (144.35)    \\
\hline
$R^2$                         & 0.11       & 0.24      & 0.33       & 0.35       & 0.19       & 0.34       & 0.35       & 0.39        \\
$N$                            & 139705 & 288827 & 38938 & 47141 & 146574 & 158479 & 88300 & 137437 \\
\hline
\multicolumn{9}{l}{Standard errors in parentheses. * $p<.1$, ** $p<.05$, ***$p<.01$}\\


\\\\
\multicolumn{9}{l}{\fontsize{13}{13}\selectfont{\textbf{\ \ (B)} Group \textbf{age} diversity}}\\
\hline\smallskip
&\multicolumn{8}{c}{Citation Count, $c^G_5$}
\\\cline{2-9}
                             &    Engineering \&  &  Health \& Medical    &    Business, Economics     &    Humanities, Literature     &    Physics \&     &    Social     &    Chemical \&     &    Life Sciences \&      \\
                             &    Computer Science     & Sciences    &   \& Management     &    \& Arts     &    Mathematics     &    Sciences     &    Material Sciences     &    Earth Sciences      \\                             
\midrule
\midrule
$d^G_{\mathit{age}}$                 & 0.59      & 8.45***   & 15.06***   & 19.82***  & 10.92***   & 23.23***   & 11.41***  & 11.28***  \\
       & (3.41)     & (0.71)    & (1.52)     & (2.73)    & (3.37)     & (3.38)     & (2.44)    & (1.95)     \\
University Ranking    
& -1.41***   & -1.04***  & -0.60**      & -0.51**  & -0.10      & -0.55*      & -0.34  & -1.31***   \\
 & (0.39)     & (0.08)    & (0.24)     & (0.26)    & (0.46)     & (0.29)     & (0.30)    & (0.23)          \\
Author's Prior Impact & 0.62***    & 1.24***   & 1.52***    & 1.61***    & 0.72***    & 1.51***    & 1.60***    & 1.53***     \\
                             & (0.01)     & (0.01)    & (0.01)     & (0.01)     & (0.02)     & (0.01)     & (0.01)     & (0.01)      \\
Year of Publication                      & 0.22     & 0.28***  & 0.38***  & 0.04     & 0.08     & 0.42*    & 0.14*   & 1.09***    \\
                             & (0.21)   & (0.01)    & (0.09)    & (0.07)   & (0.16)   & (0.23)   & (0.09)   & (0.11)      \\
Number of Authors & 0.18       & 0.24      & 0.17***    & -0.02     & 0.74       & 0.00       & 0.63      & 0.56***     \\
                             & (0.28)     & (0.15)    & (0.06)     & (0.76)    & (1.04)     & (0.21)     & (0.49)    & (0.12)    \\
const                    & 2221.02*** & 598.55*** & 1081.71*** & 1085.84*** & 1289.91*** & 2142.17*** & 1813.42*** & 2750.75***  \\
                             & (270.36)   & (22.94)   & (114.27)   & (124.13)   & (230.16)   & (194.14)   & (188.89)   & (144.35)    \\
\hline
$R^2$                         & 0.11       & 0.24      & 0.32       & 0.31       & 0.19       & 0.34       & 0.32       & 0.38        \\
$N$                            & 139705 & 288827 & 38938 & 47141 & 146574 & 158479 & 88300 & 137437 \\
\hline
\multicolumn{9}{l}{Standard errors in parentheses. * $p<.1$, ** $p<.05$, ***$p<.01$}\\
\end{tabular}
}
\end{center}
}
\end{table}
\FloatBarrier

{
\begin{center}
\resizebox{\textwidth}{!}{
\begin{tabular}{lcccccccc}
\\
\multicolumn{9}{l}{\fontsize{13}{13}\selectfont{\textbf{\ \ (C)} Group \textbf{gender} diversity}}\\
\hline\smallskip
&\multicolumn{8}{c}{Citation Count, $c^G_5$}
\\\cline{2-9}
                             &    Engineering \&  &  Health \& Medical    &    Business, Economics     &    Humanities, Literature     &    Physics \&     &    Social     &    Chemical \&     &    Life Sciences \&      \\
                             &    Computer Science     & Sciences    &   \& Management     &    \& Arts     &    Mathematics     &    Sciences     &    Material Sciences     &    Earth Sciences      \\                             
\midrule
\midrule 
$d^G_{gen}$                 & -6.34    & -0.93     & 0.57       & 1.54      & 1.55       & -0.24      & 6.34**    & -0.85       \\
                             & (4.48)   & (1.38)    & (1.67)     & (3.38)    & (4.41)     & (2.60)     & (2.93)    & (2.09)        \\
University Ranking    & -0.75    & -0.69***  & 0.06       & -1.72***  & -0.11      & -0.68**    & -1.11***  & -0.92***     \\
                             & (0.56)   & (0.12)    & (0.19)     & (0.41)    & (0.59)     & (0.29)     & (0.35)    & (0.29)      \\
Author's Prior Impact & 1.33***  & 1.67***   & 0.92***    & 1.53***   & 0.65***    & 1.47***    & 1.06***   & 1.61***   \\
                             & (0.02)   & (0.02)    & (0.01)     & (0.04)    & (0.03)     & (0.01)     & (0.05)    & (0.01)     \\
Year of Publication   & 0.70**     & 0.22***  & 0.34***  & 0.07     & 0.02     & 0.22     & 0.05    & 1.04***    \\
                             & (0.35)     & (0.03)    & (0.10)    & (0.08)   & (0.21)   & (0.23)   & (0.10)   & (0.15)      \\
Number of Authors & -0.13    & 0.79***   & 0.38***    & 1.44*     & 1.75       & 1.12***    & 1.13**    & 0.76***     \\
                             & (0.36)   & (0.19)    & (0.06)     & (0.78)    & (1.27)     & (0.19)     & (0.51)    & (0.13)  \\
const & 946.57** & 541.77*** & 2617.85*** & 468.14*** & 1579.15*** & 2669.66*** & 784.17*** & 2787.59***  \\
                             & (409.64) & (41.14)   & (104.67)   & (116.18)  & (304.95)   & (235.49)   & (133.25)  & (183.41)\\
\hline
$R^2$                         & 0.16     & 0.29      & 0.32       & 0.31      & 0.17       & 0.39       & 0.26      & 0.41       \\
$N$                            & 58288 & 188249 & 14904 & 8911 & 36949 & 30420 & 50887 & 71630 \\
\hline
\multicolumn{9}{l}{Standard errors in parentheses. * $p<.1$, ** $p<.05$, ***$p<.01$}\\


\\\\
\multicolumn{9}{l}{\fontsize{13}{13}\selectfont{\textbf{\ \ (D)} Group \textbf{affiliation} diversity}}\\
\hline\smallskip
&\multicolumn{8}{c}{Citation Count, $c^G_5$}
\\\cline{2-9}
                             &    Engineering \&  &  Health \& Medical    &    Business, Economics     &    Humanities, Literature     &    Physics \&     &    Social     &    Chemical \&     &    Life Sciences \&      \\
                             &    Computer Science     & Sciences    &   \& Management     &    \& Arts     &    Mathematics     &    Sciences     &    Material Sciences     &    Earth Sciences      \\                             
\midrule
\midrule
$d^G_{\mathit{aff}}$                 & -2.85      & 2.93***   & 2.45**     & 0.85      & 9.88***    & 5.77***    & 0.43      & 3.89***    \\
                             & (2.35)     & (0.60)    & (0.97)     & (2.70)    & (3.35)     & (1.97)     & (2.26)    & (1.36)        \\
University Ranking   & -1.35***   & -1.16***  & -0.12      & -1.29***  & -0.26      & -0.59**    & -0.79***  & -1.42***    \\
                             & (0.39)     & (0.08)    & (0.18)     & (0.36)    & (0.46)     & (0.30)     & (0.30)    & (0.24)     \\
Author's Prior Impact & 0.62***    & 1.23***   & 0.92***    & 1.49***   & 0.72***    & 1.60***    & 1.04***   & 1.53***    \\
                             & (0.01)     & (0.01)    & (0.01)     & (0.03)    & (0.02)     & (0.01)     & (0.04)    & (0.01)   \\
Year of Publication                   & 0.14     & 0.25***  & 0.28***  & 0.13*    & 0.10     & 0.58**     & 0.06    & 1.04***    \\
                             & (0.21)   & (0.01)    & (0.09)    & (0.07)   & (0.16)   & (0.23)     & (0.09)   & (0.11)      \\
Number of Authors & 0.26       & 0.55***   & 0.35***    & 1.59**    & 0.71       & 0.31       & 1.24**    & 0.64***       \\
                             & (0.28)     & (0.15)    & (0.06)     & (0.77)    & (1.05)     & (0.21)     & (0.49)    & (0.12)   \\
const                        & 2240.33*** & 622.76*** & 2370.64*** & 327.82*** & 1336.28*** & 2319.30*** & 793.64*** & 2721.82*** \\
                             & (275.40)   & (23.59)   & (91.50)    & (97.70)   & (230.77)   & (231.07)   & (117.89)  & (144.24)  \\
\hline
$R^2$                        & 0.11       & 0.24      & 0.32       & 0.30      & 0.20       & 0.35       & 0.25      & 0.39      \\
$N$                           & 38236      & 35925     & 4736      & 2738      & 61898       & 6431      & 25656      & 32279  \\
\hline
\multicolumn{9}{l}{Standard errors in parentheses. * $p<.1$, ** $p<.05$, ***$p<.01$}\\

\\\\
\multicolumn{9}{l}{\fontsize{13}{13}\selectfont{\textbf{\ \ (E)} Group \textbf{discipline} diversity}}\\
\hline\smallskip
&\multicolumn{8}{c}{Citation Count, $c^G_5$}
\\\cline{2-9}
                             &    Engineering \&  &  Health \& Medical    &    Business, Economics     &    Humanities, Literature     &    Physics \&     &    Social     &    Chemical \&     &    Life Sciences \&      \\
                             &    Computer Science     & Sciences    &   \& Management     &    \& Arts     &    Mathematics     &    Sciences     &    Material Sciences     &    Earth Sciences      \\                             
\midrule
\midrule
$d^G_{dsp}$                    & 7.39    & 15.08***  & 6.92      & 31.35*** & 24.35*** & 7.00        & 25.05*** & 15.77***    \\
                             & (9.91)   & (1.66)    & (5.47)    & (6.68)   & (7.37)    & (13.70)     & (7.08)   & (3.42)      \\
University Ranking                 & -2.46*** & -1.01***  & -0.49     & -1.36*** & -0.96     & 0.28        & -0.85*   & -1.75***    \\
                             & (0.55)   & (0.10)    & (0.30)    & (0.51)   & (0.64)    & (0.53)      & (0.48)   & (0.32)      \\
Author's Prior Impact & 0.62***  & 1.35***   & 0.91***   & 1.45***  & 0.69***   & 1.80***     & 0.96***  & 1.55***     \\
                             & (0.01)   & (0.01)    & (0.01)    & (0.04)   & (0.03)    & (0.02)      & (0.05)   & (0.01)      \\
Year of Publication                   & 0.15     & 0.28***  & 0.29***  & 0.01    & -0.01     & 0.71***     & 0.19**  & 1.13***    \\
                             & (0.22)   & (0.02)    & (0.09)    & (0.08)   & (0.18)    & (0.25)      & (0.10)   & (0.11)      \\
Number of Authors                & 0.02     & 0.05***   & 0.02***   & 0.24*    & 0.28      & 0.10***     & 0.17***  & 0.04***     \\
                             & (0.02)   & (0.02)    & (0.01)    & (0.14)   & (0.20)    & (0.03)      & (0.05)   & (0.01)      \\
const                        & -253.60  & 566.42*** & 598.47*** & 24.34    & 76.50     & -1412.96*** & 387.01** & 2278.47***  \\
                             & (446.69) & (32.93)   & (182.43)  & (161.50) & (352.32)  & (502.55)    & (190.31) & (226.61)    \\
\hline
R2                           & 0.10     & 0.25      & 0.26      & 0.29     & 0.18      & 0.35        & 0.21     & 0.38        \\
N                            & 104088    & 141917     & 20801    & 12238     & 100839     & 24773      & 65607     & 98006     \\
\hline
\multicolumn{9}{l}{Standard errors in parentheses. * $p<.1$, ** $p<.05$, ***$p<.01$}\\
\end{tabular}
}
\end{center}
}

\begin{table}[H]
{\fontsize{10}{10}\selectfont{
\caption{\textbf{Coarsened exact matching of group ethnic diversity}. $T$ and $C$ are the treatment and control populations respectively; $T'$ and $C'$ are the populations of matched treatment and matched control papers respectively; $\mathcal{L}_1$ is  the multivariate imbalance statistic~\cite{iacus2012causal}; $\delta$ is the relative impact gain of $T'$ over $C'$, i.e., $\delta = 100\times(\langle c^G_5 \rangle_{T'} - \langle c^G_5 \rangle_{C'})/\langle c^G_5 \rangle_{C'}$. A t-test shows that $\delta$ is statistically significant; see the resulting $p$-values. Since the academic impact $\langle c^G_5 \rangle$ is sensitive to extremal values, we bootstrap a 95\% confidence interval ($CI_{.95}$). Here, university ranking corresponds to the \textit{average rank} of all universities in the paper, as opposed to the \textit{highest ranked} university in the paper, which is the case in Supplementary Table~5. For more details, see Supplementary Note~6.} 
\label{tab:CEMGroup}
\begin{center}
\begin{tabular}{lcccccccccccc}
\toprule
 & $|T|$ & $|C|$ &  $|T'|$ & $|C'|$ &  $\mathcal{L}_1$ & $\delta$ & $CI_{.95}$ & $p$\\
\midrule
$T : d^G_{eth} > P_{90}(d^G_{eth})$  & 17,802 & 45,710 &  13,530 &  16,008 &  0.39 &  10.63 &  [8.10, 12.38]&  0.003\\
$C : d^G_{eth} \leq P_{10}(d^G_{eth})$\\
\midrule
$T: d^G_{eth} > P_{80}(d^G_{eth})$  & 24,827 & 45,710 &  18,965 &  16,165 &  0.38 &  10.22  &  [8.12, 12.02] &  0.0009\\
$C: d^G_{eth} \leq P_{20}(d^G_{eth})$\\
\midrule
$T: d^G_{eth} > P_{70}(d^G_{eth})$  & 56,662 & 58,889 &  51,782 &  39,216 &  0.27 &  4.93 &  [3.74, 5.97] &  0.008\\
$C: d^G_{eth} \leq P_{30}(d^G_{eth})$\\
\midrule

$T: d^G_{eth} > P_{60}(d^G_{eth})$ & 63,129 & 78,340 &  57,279 &  58,199 &  0.29 &  5.14 &   [4.12, 6.17] &  0.003\\
$C: d^G_{eth} \leq P_{40}(d^G_{eth})$\\
\midrule
$T: d^G_{eth} > P_{50}(d^G_{eth})$  & 63,129 & 127,629 &  58,292 &  70,627 &  0.27 &  3.37 &  [2.45, 4.25] &  0.018\\
$C: d^G_{eth} \leq P_{50}(d^G_{eth})$\\
\midrule
\end{tabular}
\end{center}
}}
\end{table}

\begin{table}[H]
{\fontsize{10}{10}\selectfont{
\caption{\textbf{Coarsened exact matching of individual ethnic diversity}. The notation is as per Table~\ref{tab:CEMGroup}.} 
\label{tab:CEMIndividual}
\begin{center}
\begin{tabular}{lcccccccccccc}
\toprule
& $|T|$ & $|C|$ &  $|T'|$ & $|C'|$ &  $\mathcal{L}_1$ & $\delta$ &  $CI_{.95}$ & $p$\\
\midrule
$T : d^I_{eth} > P_{90}(d^I_{eth})$ & 113,883 & 68,563 & 16,512 & 20,599  & 0.47 & 47.67 & [44.49, 49.92] &  2.04e-39\\
$C : d^I_{eth} \leq P_{10}(d^I_{eth})$\\
\midrule

$T: d^I_{eth} > P_{80}(d^I_{eth})$ 
& 139,015 & 136,837 & 65,412 & 50,240 & 0.35 & 43.54 & [42.61, 45.05] & 1.50e-156\\
$C: d^I_{eth} \leq P_{20}(d^I_{eth})$\\
\midrule

$T: d^I_{eth} > P_{70}(d^I_{eth})$ 
& 223,747 & 205,686 & 128,001 & 117,560 & 0.32 & 28.75& [28.10, 29.46] & 1.65e-211\\
$C: d^I_{eth} \leq P_{30}(d^I_{eth})$\\
\midrule

$T: d^I_{eth} > P_{60}(d^I_{eth})$ & 280,514 & 274,209 & 184,749 & 143,683 & 0.29 & 23.86 & [22.86, 23.98] & 5.96e-218 \\
$C: d^I_{eth} \leq P_{40}(d^I_{eth})$\\
\midrule

$T: d^I_{eth} > P_{50}(d^I_{eth})$ & 356,564 & 329,066 & 242,123 & 240,237 & 0.28 & 15.77 & [15.21, 15.95] & 3.23e-158\\
$C: d^I_{eth} \leq P_{50}(d^I_{eth})$\\
\midrule
\end{tabular}
\end{center}
}}
\end{table}

\end{document}



\baselineskip24pt

\date{\vspace{-0.1ex}}
\maketitle 
\ \\
\ \\

\vspace{-1cm}

{\noindent\Large \textbf{Supplementary Figures}}

\begin{figure}[H]
\centering
\hspace*{-1.4cm} \includegraphics[width=1.2\textwidth]{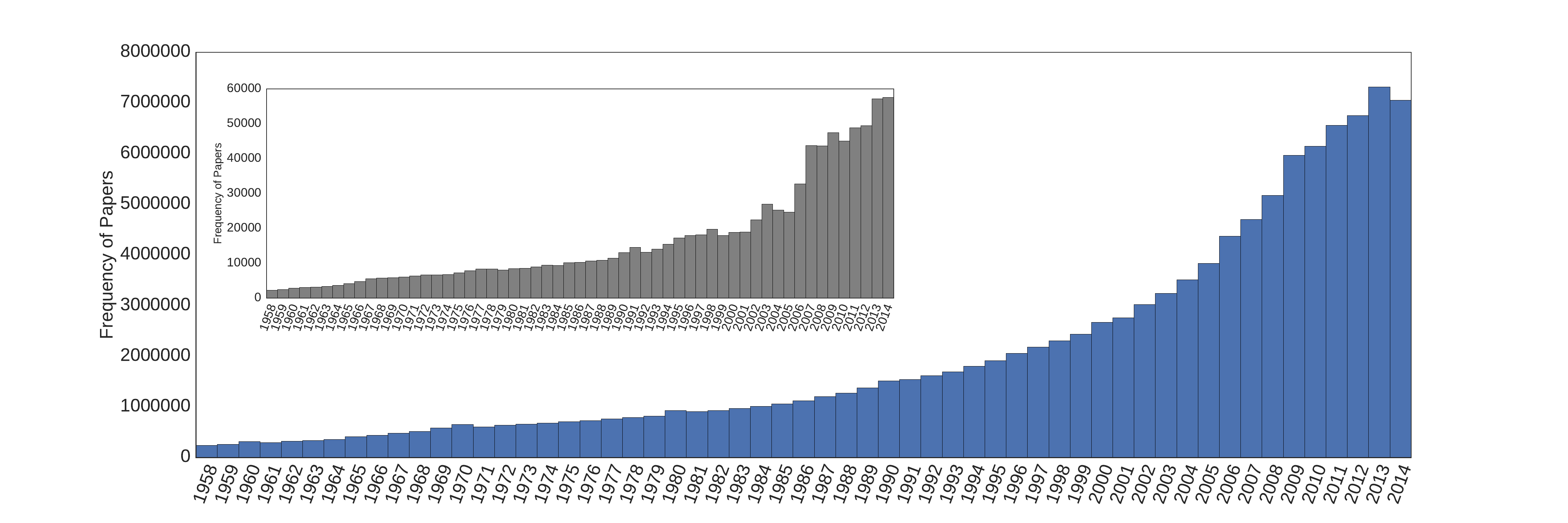}
\caption{The distribution of papers published each year in the Microsoft Academic Graph corpus. The inset shows the distribution of our sample set, consisting of 1,045,401 papers.}
\label{fig:PapersHist}
\end{figure}

\newpage

\begin{figure}[H]
\centering
\includegraphics[width=0.7\textwidth]{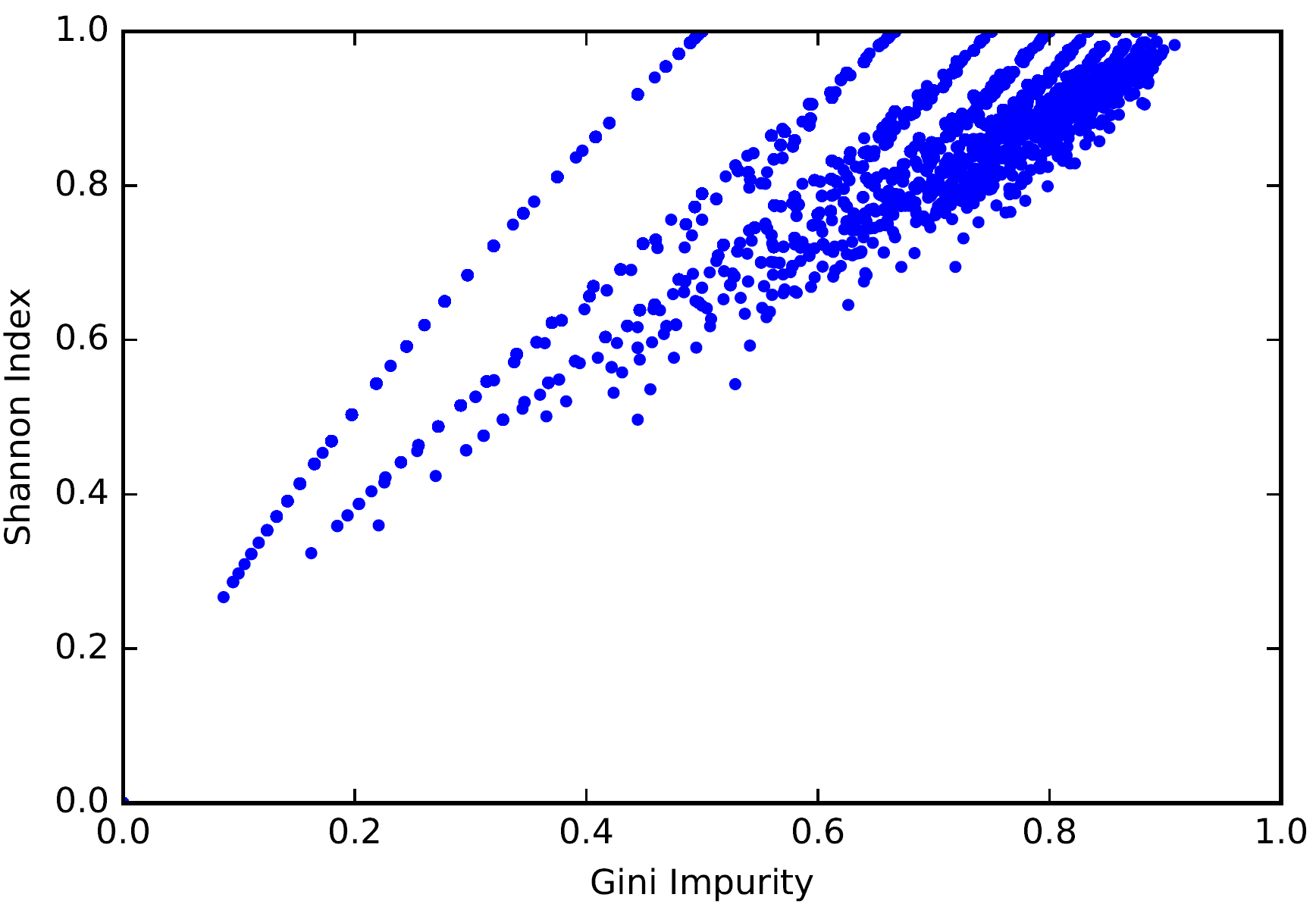}
\caption{For every group of scientists that coauthored a paper in the Microsoft Academic Graph dataset, we measured the ethnic diversity using both the Shannon entropy and the Gini index. The two are plotted against each other, showing a clear correlation (Pearson's $r=0.93$ and $p<0.0001$).} 
\label{fig:ShannonVsGiniImpurity}
\end{figure}

\newpage

\begin{figure}[H]%
\centering
\includegraphics[height=18.3cm,keepaspectratio]{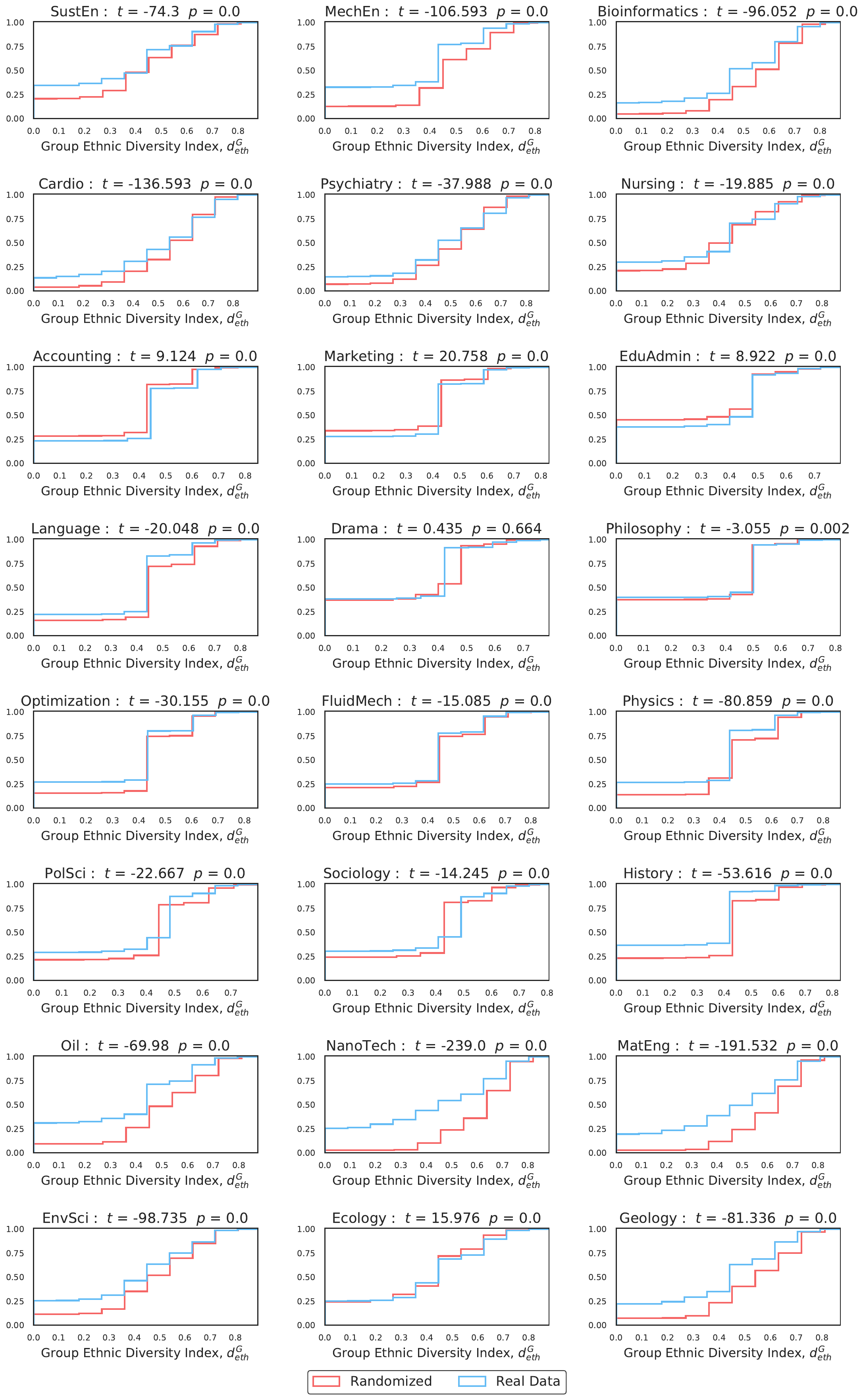}
\caption{Cumulative distribution function (CDF) of group \textbf{ethnic diversity}, $d^G_{\mathit{eth}}$, for the real and randomized data. In all 24 subfields, groups with low $d^G_{\mathit{eth}}$ are more common in reality than expected by random chance, highlighting the fact that homophily does indeed exist in academia. For all subfields, the difference between the two datasets is statistically significant ($p \ll .05$).}
\label{fig:Surrogate_EthnicIndex}
\end{figure}

\newpage

\begin{figure}[H]%
\centering
\includegraphics[height=18.3cm,keepaspectratio]{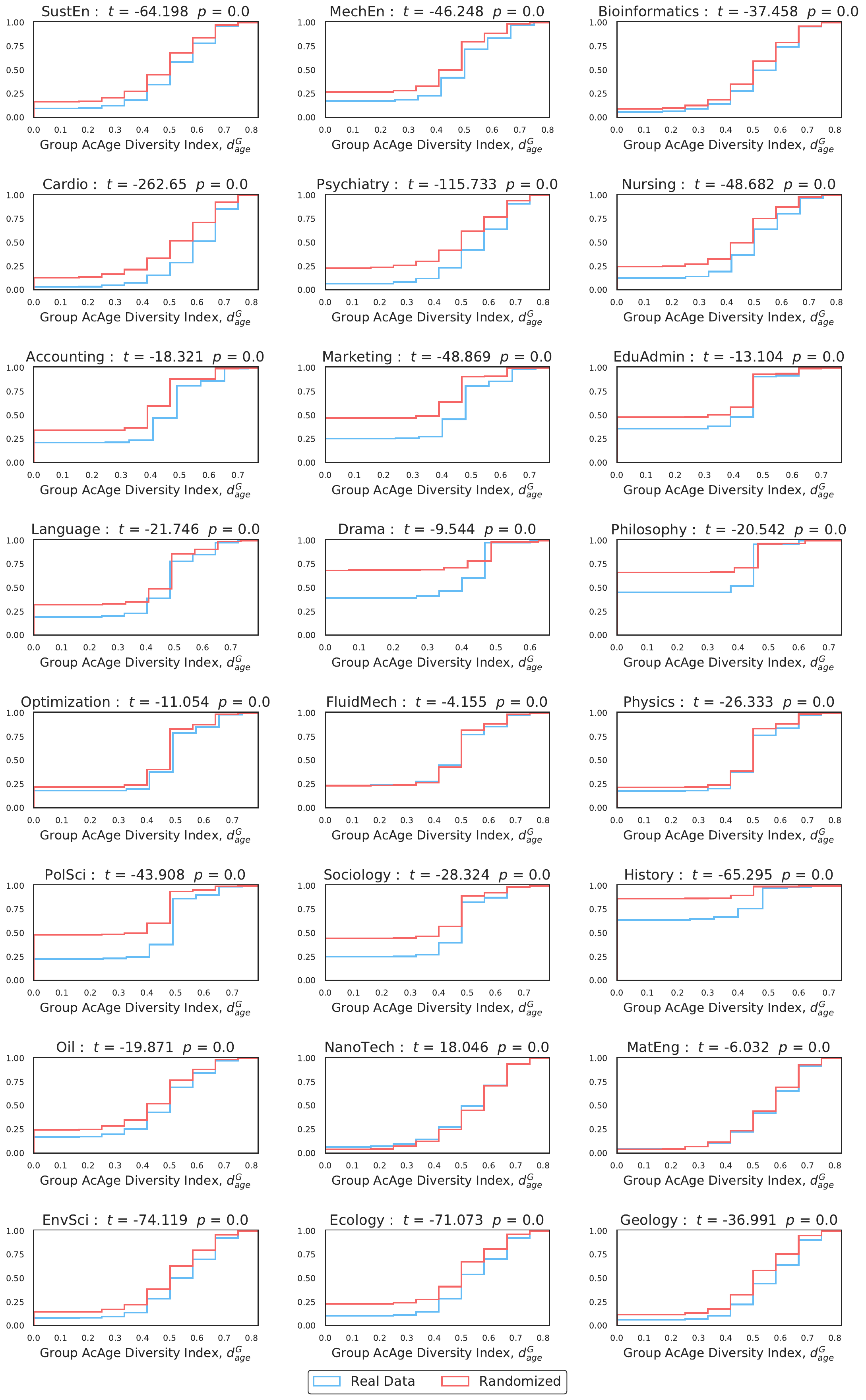}
\caption{Cumulative distribution function (CDF) of group \textbf{age diversity}, $d^G_{\mathit{age}}$, for the real and randomized data. In all 24 subfields, groups with low $d^G_{\mathit{age}}$ are more common in reality than expected by random chance, highlighting the fact that homophily does indeed exist in academia. For all subfields, the difference between the two datasets is statistically significant ($p \ll .05$).}
\label{fig:Surrogate_AcAgeIndex}
\end{figure}

\newpage

\newpage

\begin{figure}[H]%
\centering
\includegraphics[height=18.3cm,keepaspectratio]{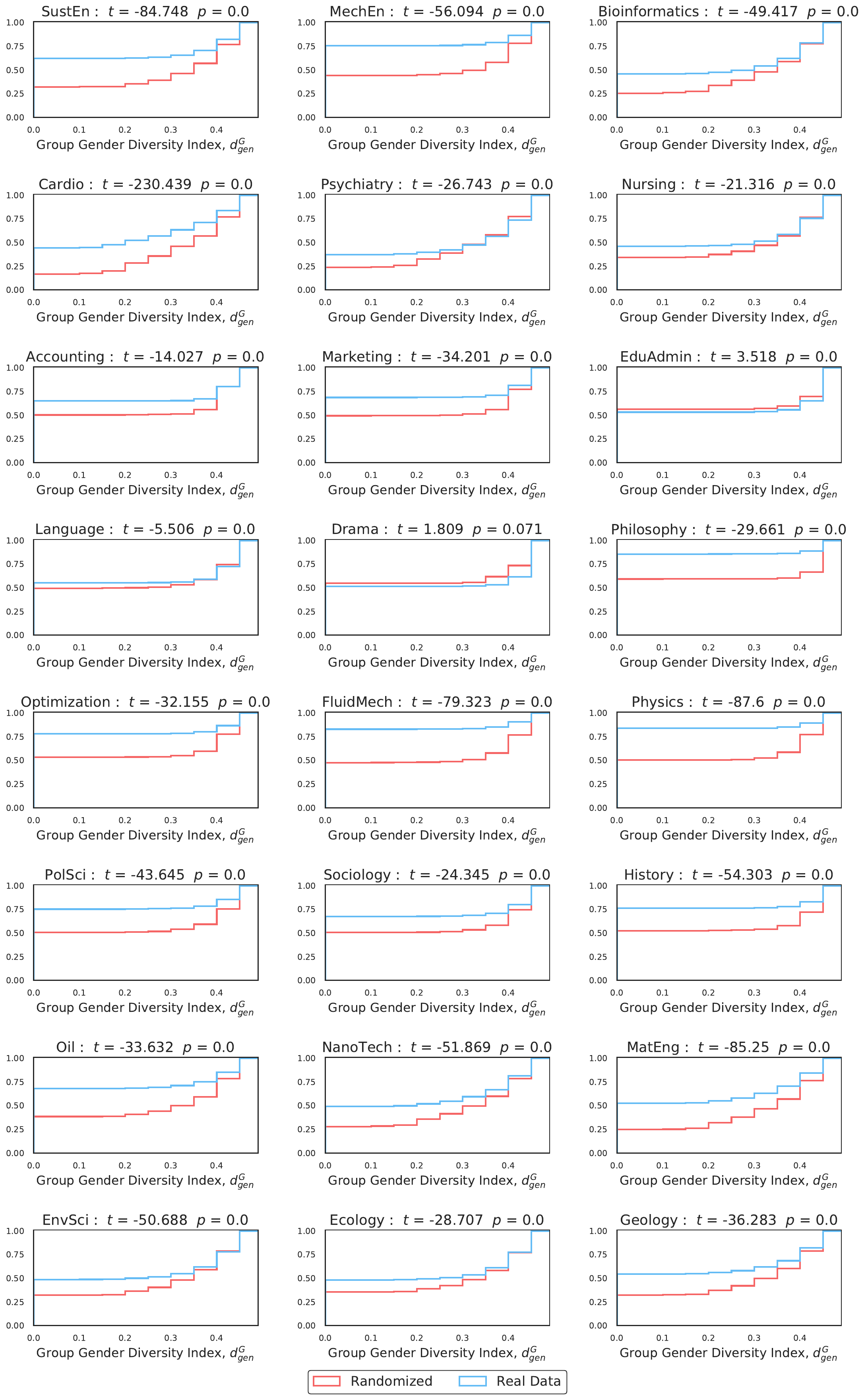}
\caption{Cumulative distribution function (CDF) of group \textbf{gender diversity}, $d^G_{\mathit{gen}}$, for the real and randomized data. In all 24 subfields, groups with low $d^G_{\mathit{gen}}$ are more common in reality than expected by random chance, highlighting the fact that homophily does indeed exist in academia. For all subfields, the difference between the two datasets is statistically significant ($p \ll .05$).}
\label{fig:Surrogate_GenderIndex}
\end{figure}

\newpage

\begin{figure}[H]%
\centering
\includegraphics[height=18.3cm,keepaspectratio]{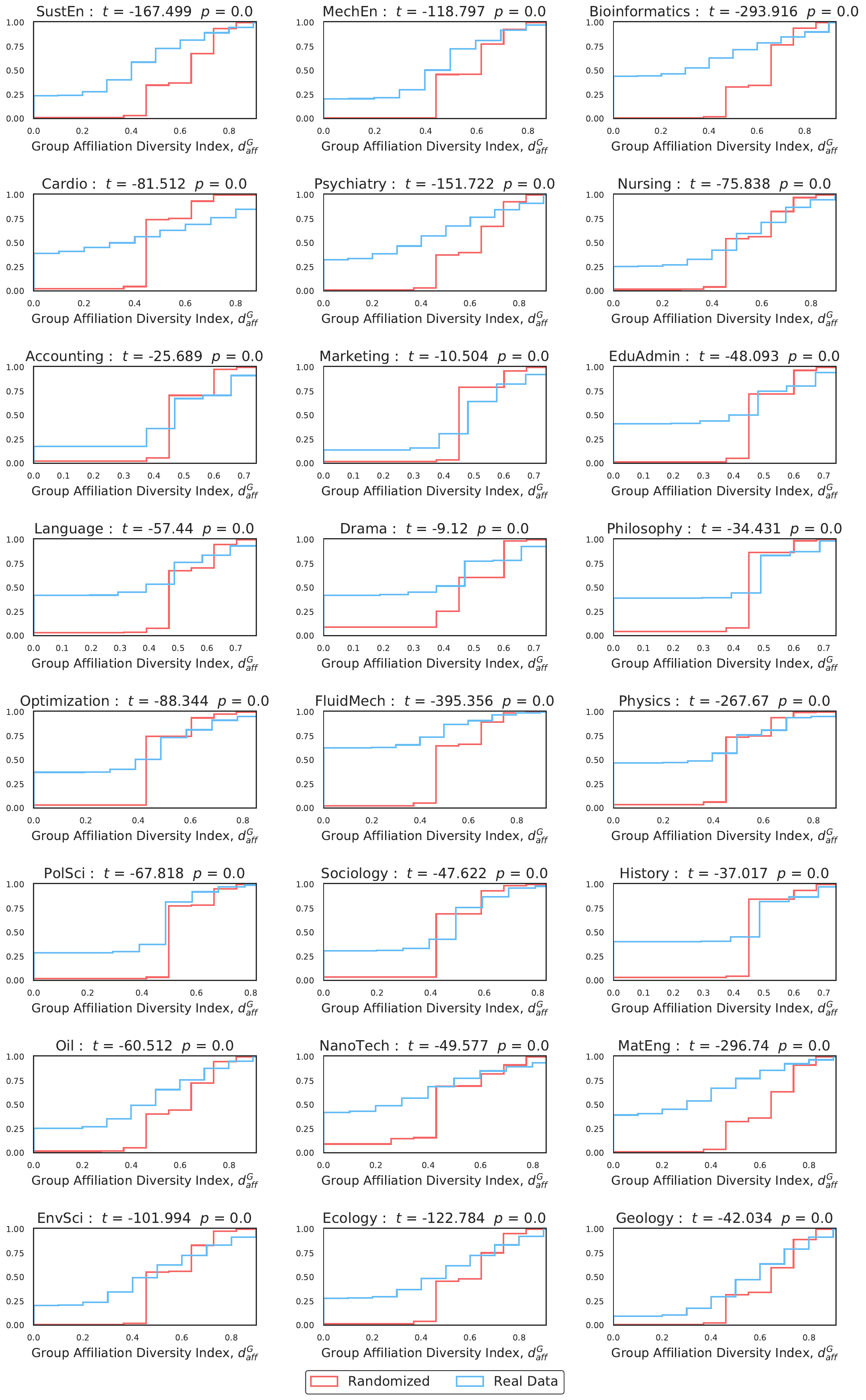}
\caption{Cumulative distribution function (CDF) of group \textbf{affiliation diversity}, $d^G_{\mathit{aff}}$, for the real and randomized data. In all 24 subfields, groups with low $d^G_{\mathit{aff}}$ are more common in reality than expected by random chance, highlighting the fact that homophily does indeed exist in academia. For all subfields, the difference between the two datasets is statistically significant ($p \ll .05$).}
\label{fig:Surrogate_AffIndex}
\end{figure}

\newpage

\begin{figure}[H]
\centering
\includegraphics[width=0.7\textwidth]{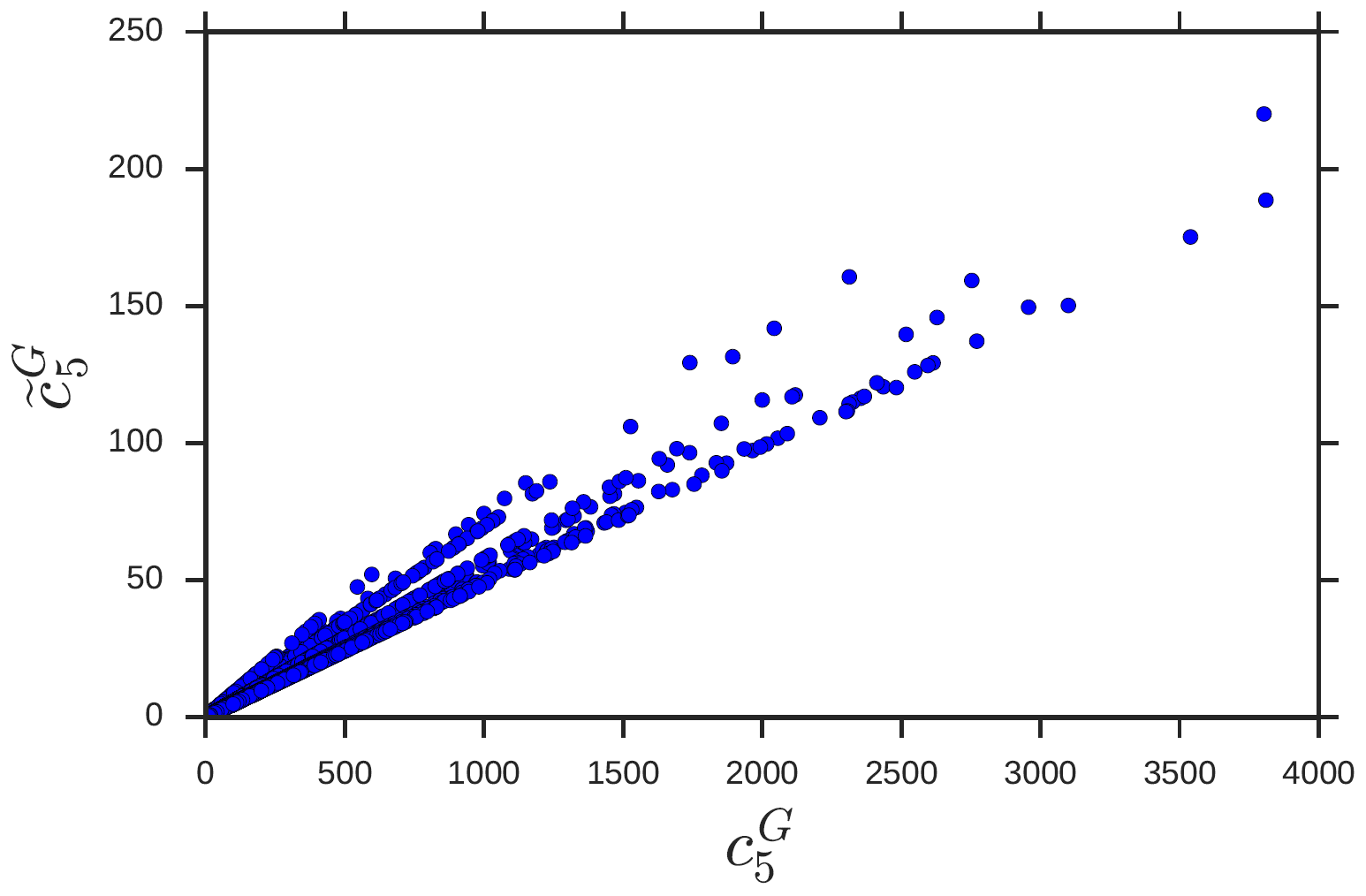}
\caption{Using 500,000 papers sampled from the entire MAG dataset, we compare $c^G_5$ with an alternative, normalized measure of impact, denoted as $\widetilde{c}^G_5$ (see Supplementary Note \ref{SM:CitationsCount} for more details). As can be seen, the two are very strongly correlated (over the entire dataset, Pearson's $r=0.965$ and $p<0.0001$).}
\label{fig:c5_vs_normalized_c5}
\end{figure}

\clearpage

\begin{figure}[H]%
\centering
\includegraphics[height=16.5cm,keepaspectratio]{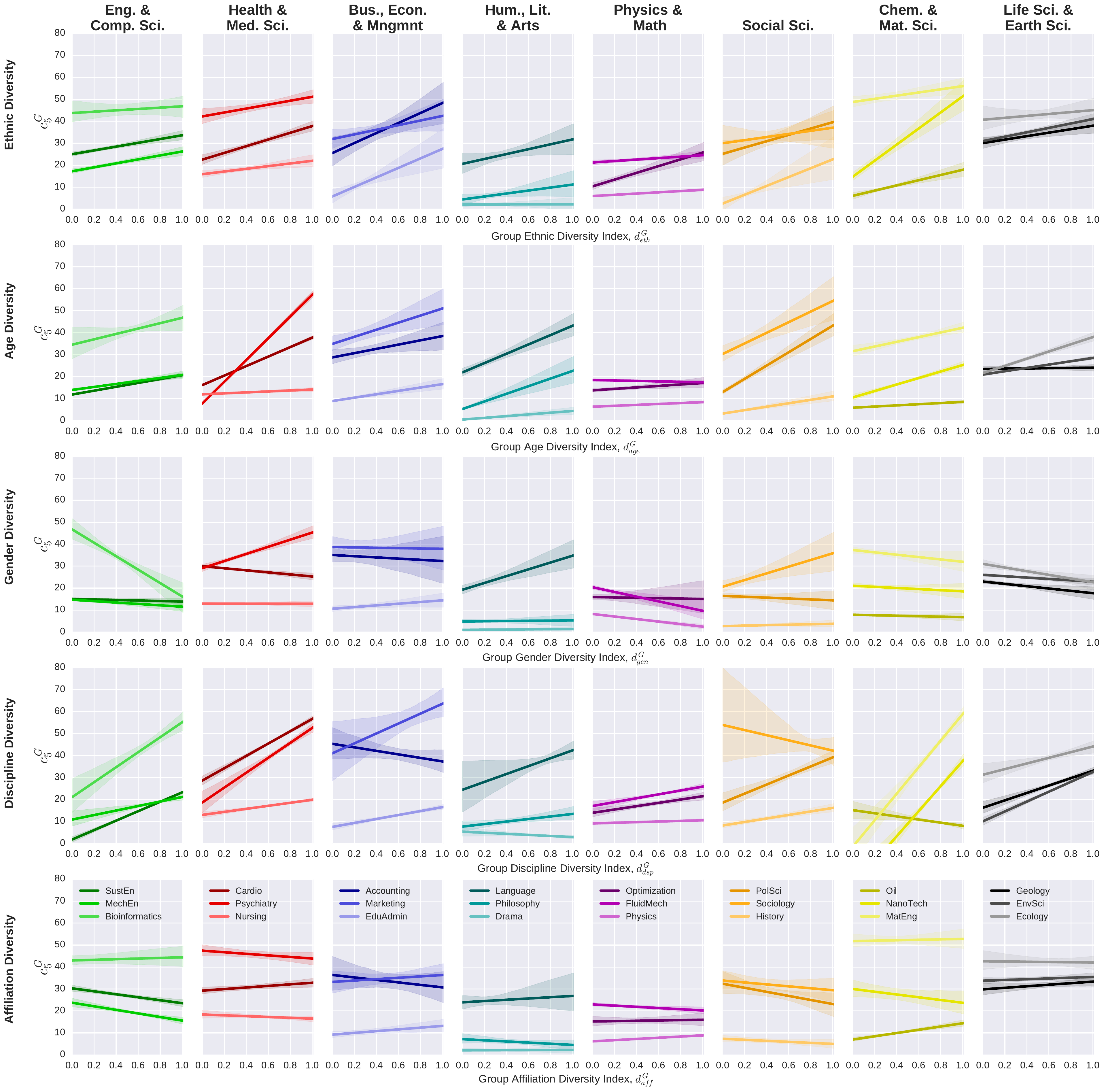}
 \caption{Group diversity against $c^G_5$ in each of the 24 subfields, which are grouped into the 8 main fields in Google Scholar. 
In the case of group \textit{ethnic} diversity and group \textit{age} diversity, every significant correlation with $c^G_5$ is positive, and nearly all correlations were significant. This, however, does not hold for the remaining group diversity indices (see the corresponding p-values in Supplementary Table~\ref{tab:GroupDiversityIndexCorr}).}
     
\label{fig:paperindices_subfields}
\end{figure}

\newpage

\begin{figure}[H]%
\centering
\includegraphics[width=1\textwidth]{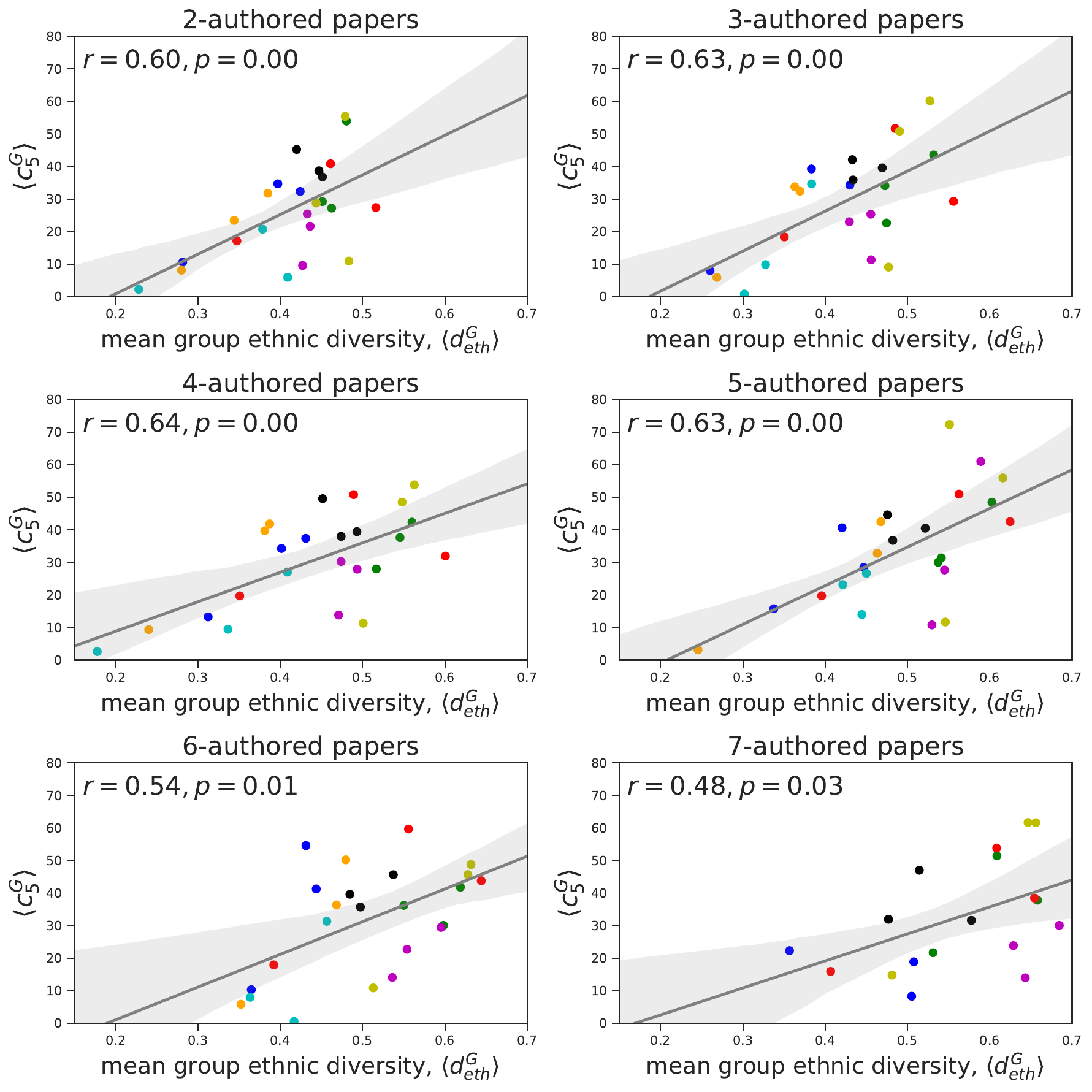}
 \caption{Mean group ethnic diversity against mean paper impact in each subfield while controlling for the number of authors. In each subplot, the color indicates the main field, while the solid line and the shaded area represent the regression line and the 95\% confidence interval, respectively. Each regression has also been annotated with the corresponding Pearson's $r$ and $p$ values. For each subfield, the subplots depict the mean group ethnic diversity, $\langle d^G_{\mathit{eth}}\rangle$, against the mean five-year citation count, $\langle c_5^G\rangle$, taken over papers in that subfield while controlling for the number of authors in each paper. 
}
\label{fig:groupindices_subfields:NumOfAuthors}
\end{figure}

\newpage

\begin{figure}[H]%
\centering
\includegraphics[height=16.5cm,keepaspectratio]{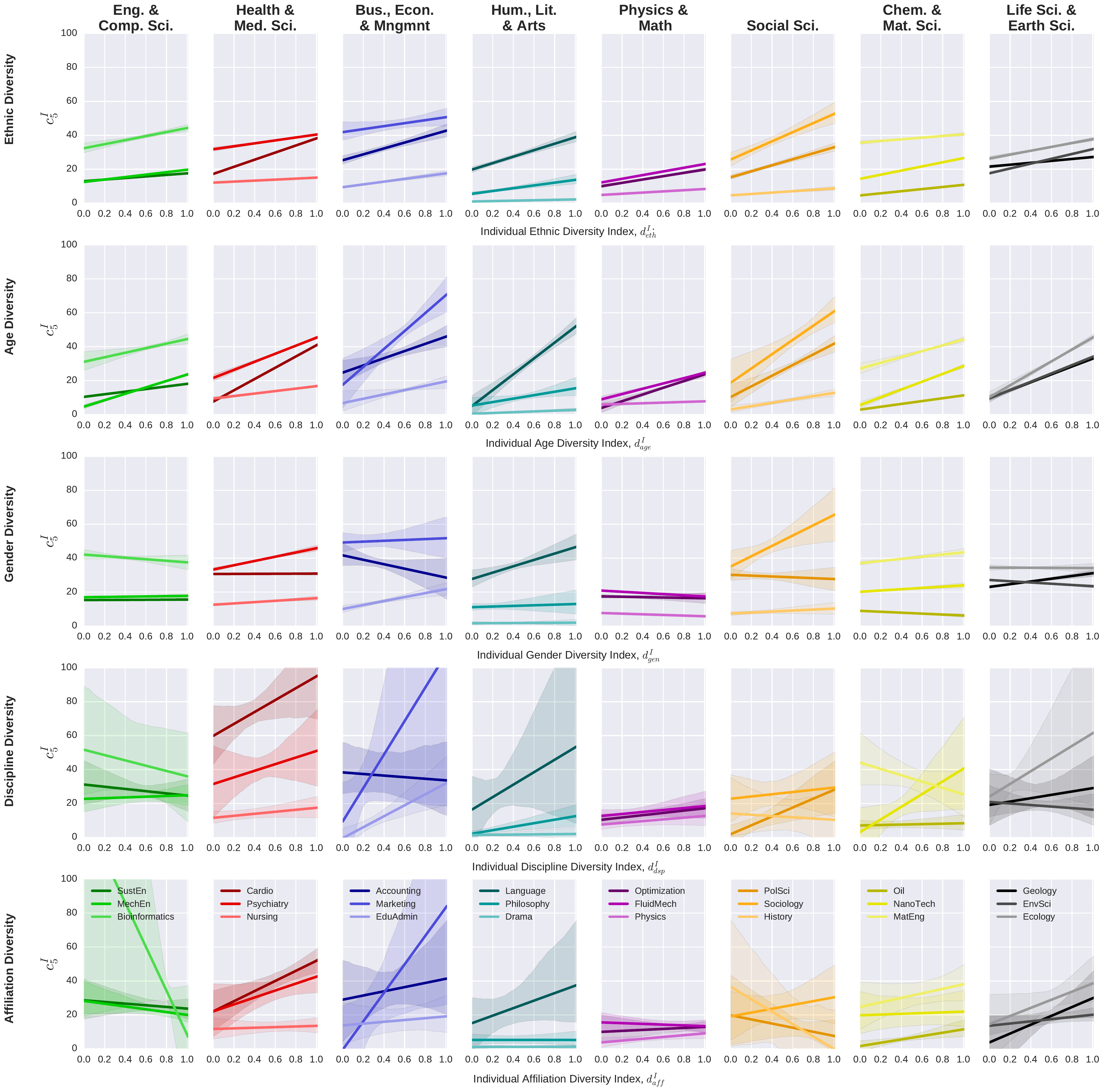}
  \caption{Individual diversity against $c^I_5$ in each of the 24 subfields, which are grouped into the 8 main fields in Google Scholar. 
In the case of individual \textit{ethnic} diversity, every correlation with $c^I_5$ is significantly positive. The same holds for individual \textit{age} diversity, with the exception of two subfields (Philosophy and Drama) for which the correlations are positive but not significant. This, however, does not hold for the remaining individual diversity indices (see the corresponding p-values in Supplementary Table~\ref{tab:IndividualDiversityIndexCorr}).}
\label{fig:authorindices_subfields}
\end{figure}

\newpage
\begin{figure}[H]%
\centering
\includegraphics[width=.77\textwidth]{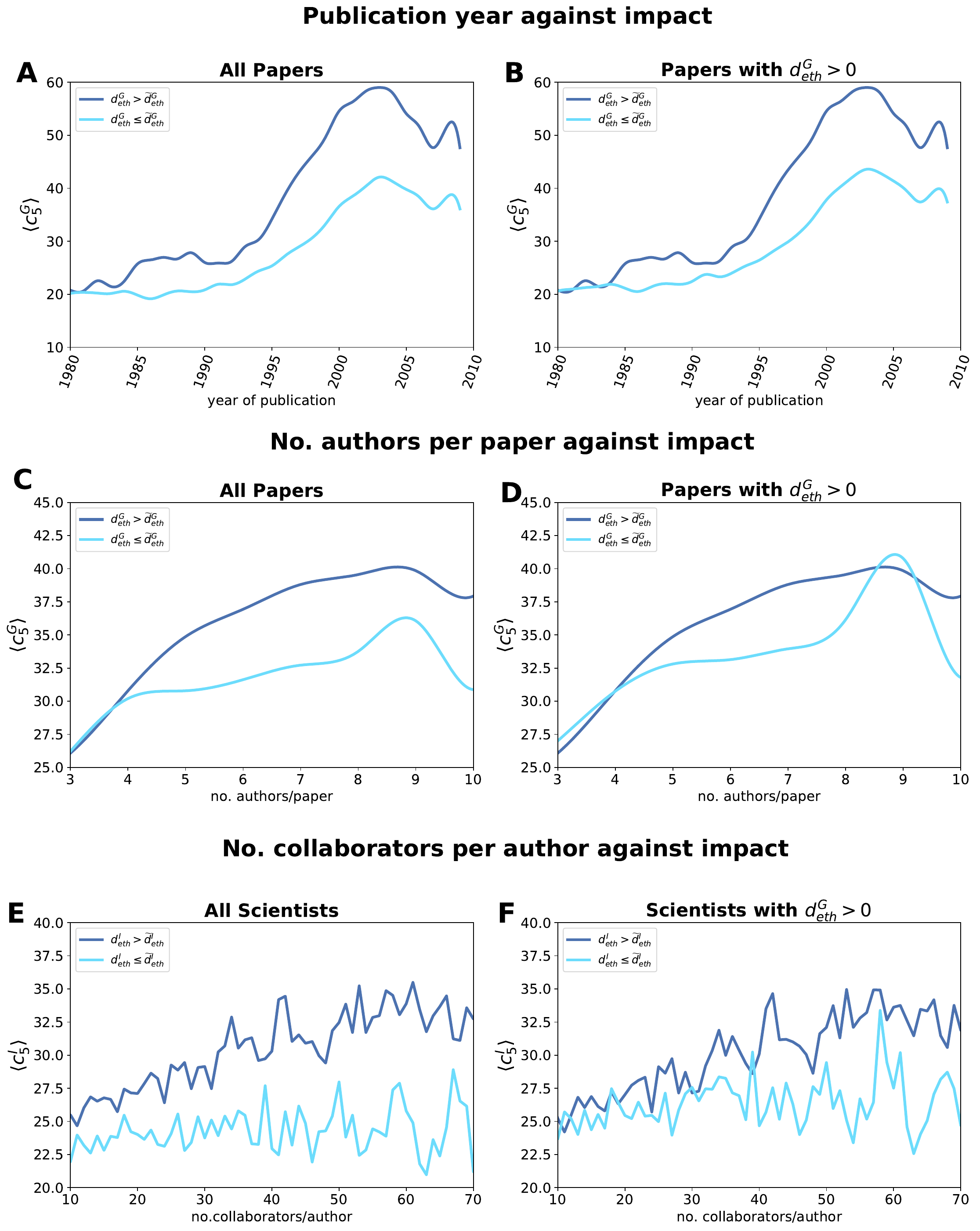}
 \caption{ {Excluding papers and scientists with no ethnic diversity.}
 \textbf{(A)} Comparison between ``diverse'' papers (where $d^G_{\mathit{eth}} > \widetilde{d}^G_{\mathit{eth}}$) against ``non-diverse'' ones (where $d^G_{\mathit{eth}} \leq \widetilde{d}^G_{\mathit{eth}}$) given different publication years.
 \textbf{(B)} Same analysis as in (A), but after excluding all papers for which $d^G_{\mathit{eth}}=0$.
 \textbf{(C)} Comparison between diverse and non-diverse papers given different numbers of authors.
 \textbf{(D)} Same analysis as in (C), but after excluding all papers for which $d^G_{\mathit{eth}}=0$.
 \textbf{(E)} Comparison between diverse scientists (whose $d^I_{\mathit{eth}} > \widetilde{d}^I_{\mathit{eth}}$) against non-diverse ones (whose $d^I_{\mathit{eth}} \leq \widetilde{d}^I_{\mathit{eth}}$) given different numbers of collaborators.
 \textbf{(F)} Same analysis as in (E), but after excluding all scientists for which $d^I_{\mathit{eth}}=0$.}
\label{fig:excludingZeroDiversity}
\end{figure}

\begin{figure}[H]%
\centering
\includegraphics[height=16.6cm]{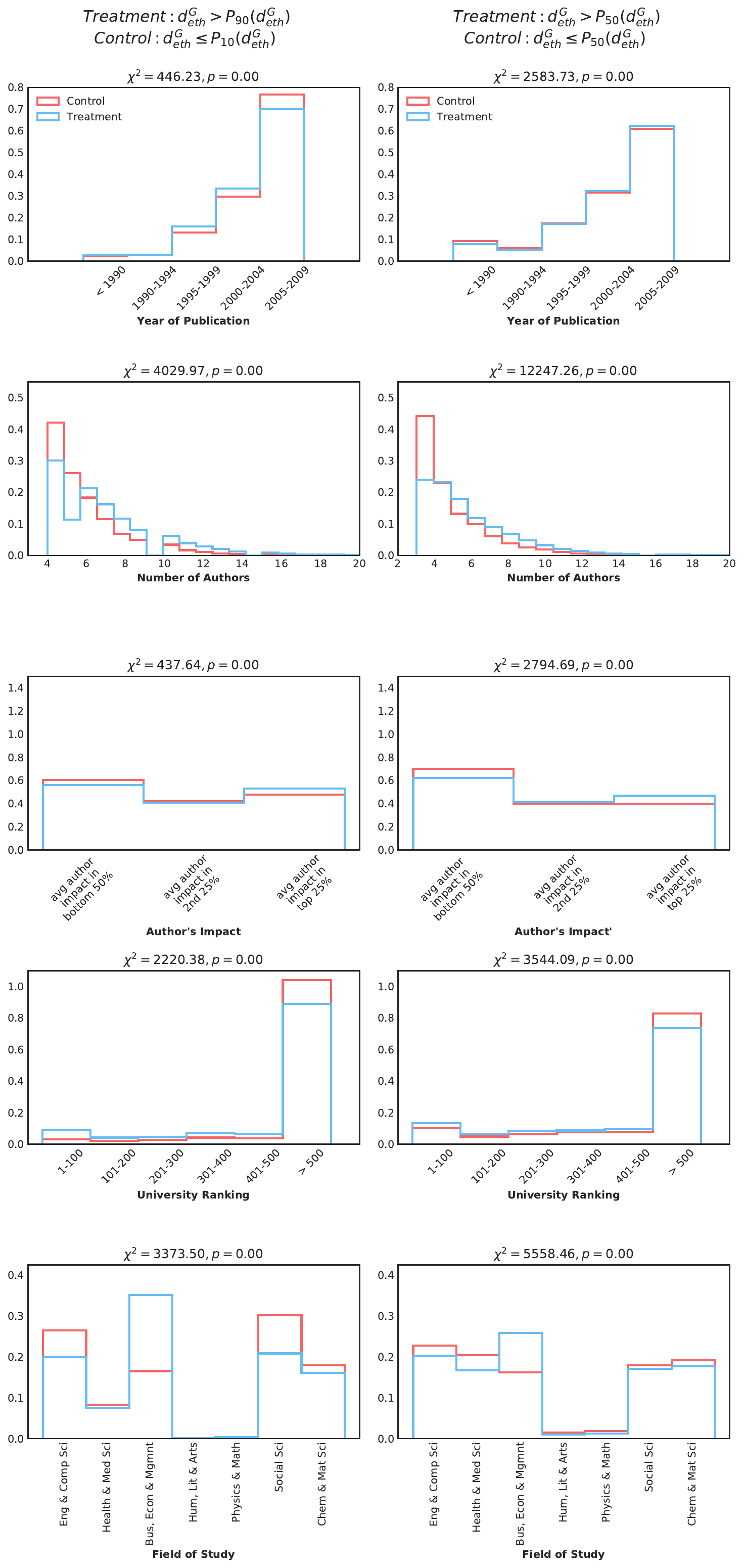}
 \caption{ The distribution (before matching) of the treatment and control groups for each confounding factor in the coarsened exact matching (CEM) when studying the causal effect of \textbf{group} ethnic diversity on scientific impact. The treatment and the control sets consist of papers for which $d^G_{\mathit{eth}} > P_{100-i}\left(d^G_{\mathit{eth}}\right)$ and $d^G_{\mathit{eth}} \leq P_i\left(d^G_{\mathit{eth}}\right)$, respectively, where $P_i\left(d^G_{\mathit{eth}}\right)$ denotes the $i^{th}$ percentile of $d^G_{\mathit{eth}}$. In our study, we repeated this process using $i=10, 20, 30, 40, 50$, but the figure only depicts the distributions for the cases where $i=10$ (left column) and where $i=50$ (right column). For the sub-figures in the second row (number of authors), the $x$-axes have been truncated to exclude outliers. Each subfigure has been annotated with its corresponding Chi-Squared test results, all of which were significant ($p < 0.0001$).}
\label{fig:confoundingFactorsDistributions:group}
\end{figure}

\begin{figure}[H]%
\centering
\includegraphics[height=16.6cm]{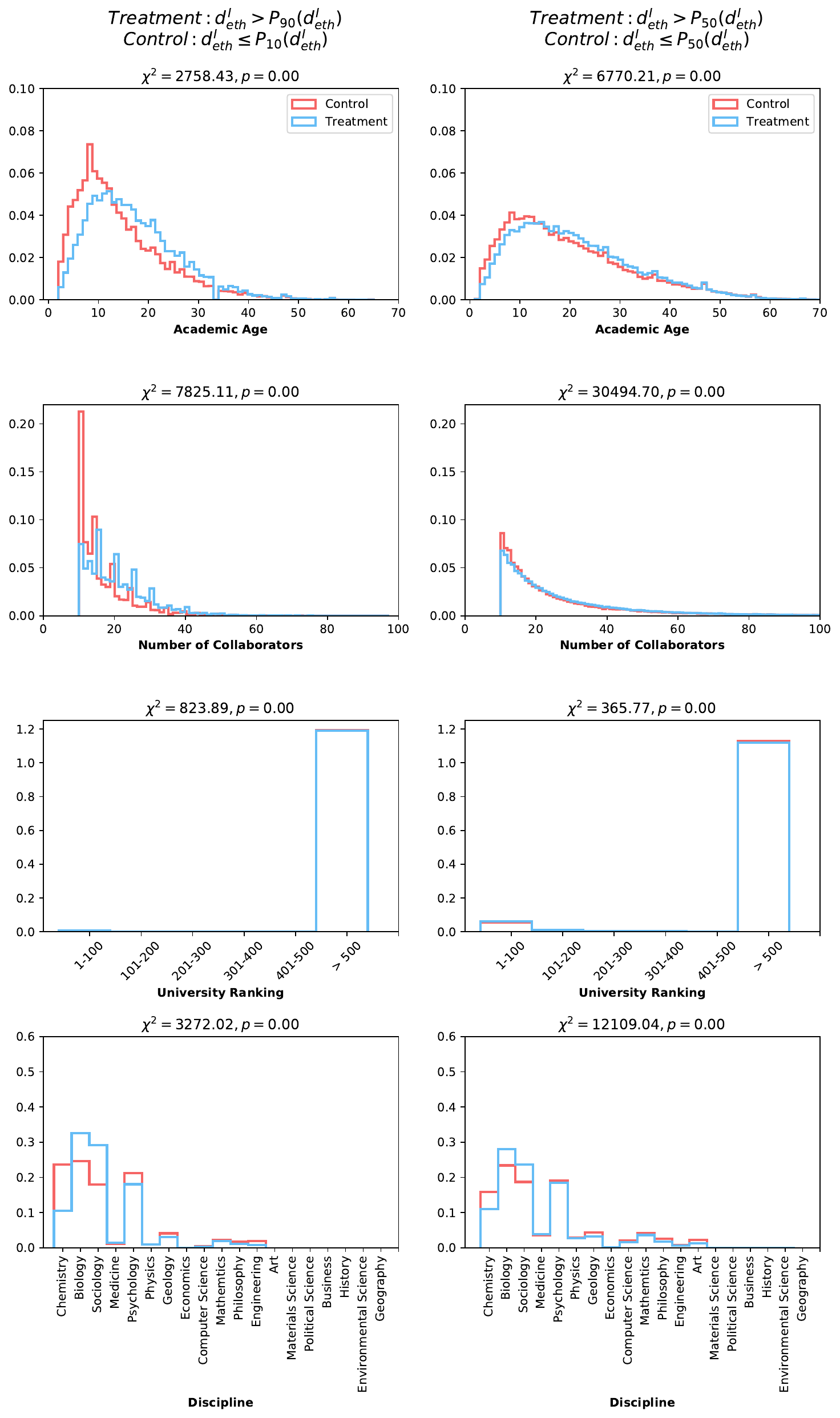}
 \caption{ The distribution (before matching) of the treatment and control groups for each confounding factor in the coarsened exact matching (CEM) when studying the causal effect of \textbf{individual} ethnic diversity on scientific impact. The treatment and the control sets consist of scientists for which $d^I_{\mathit{eth}} > P_{100-i}\left(d^I_{\mathit{eth}}\right)$ and $d^I_{\mathit{eth}} \leq P_i\left(d^I_{\mathit{eth}}\right)$, respectively, where $P_i\left(d^I_{\mathit{eth}}\right)$ denotes the $i^{th}$ percentile of $d^I_{\mathit{eth}}$. In our study, we repeated this process using $i=10, 20, 30, 40, 50$, but the figure only depicts the distributions for the cases where $i=10$ (left column) and where $i=50$ (right column). For the sub-figures in the second row (number of collaborators), the $x$-axes have been truncated to exclude outliers. Each subfigure has been annotated with its corresponding Chi-Squared test results, all of which were significant ($p < 0.0001$).}
\label{fig:confoundingFactorsDistributions:individual}
\end{figure}

\begin{figure}[H]%
\centering
\includegraphics[height=18cm]{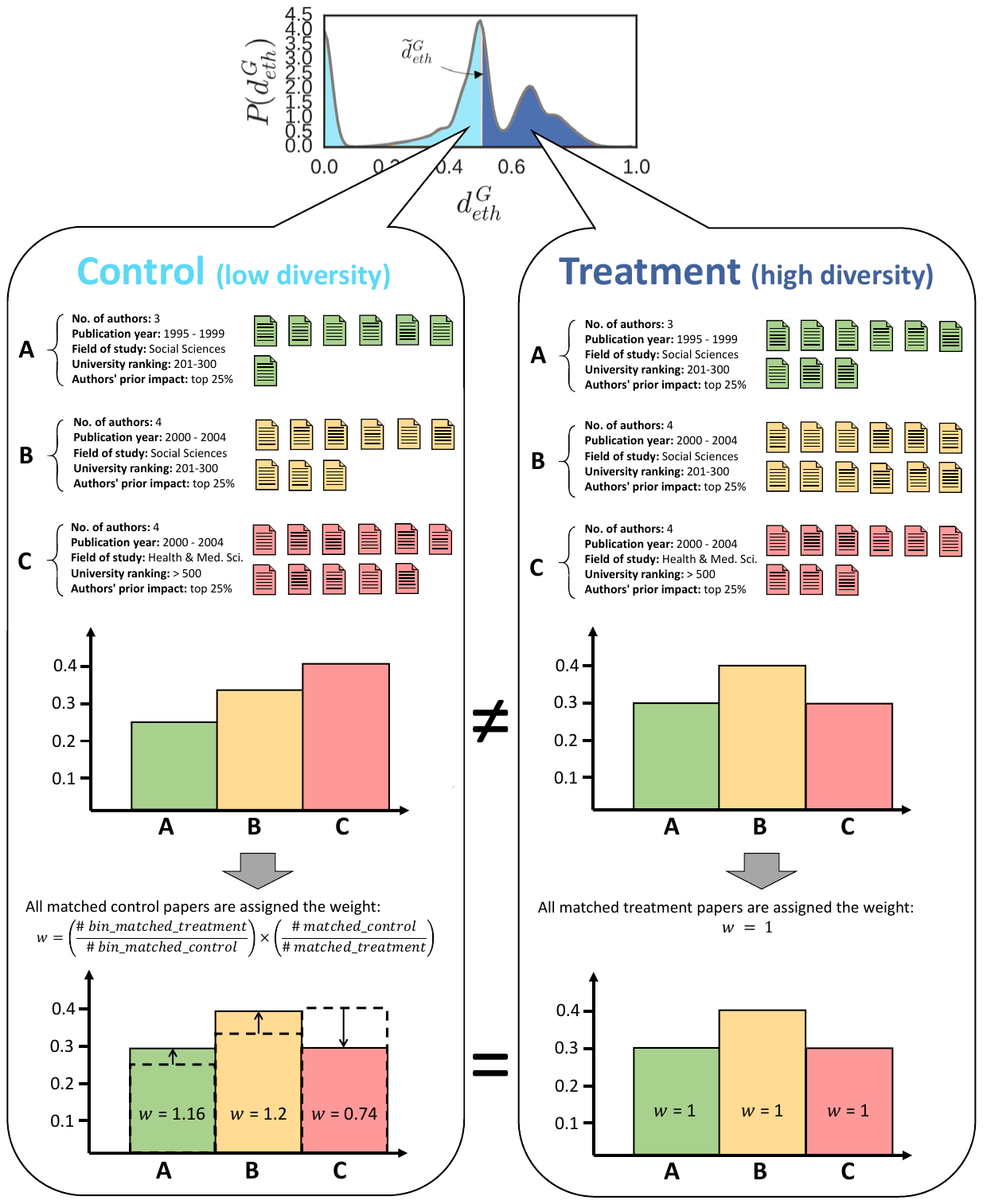}
 \caption{ {An illustration of coarsened exact matching (CEM).} 
 The treatment set consists of papers for which $d^G_{\mathit{eth}} > P_{100-i}\left(d^G_{\mathit{eth}}\right)$, and the control set of papers for which $d^G_{\mathit{eth}} \leq P_i\left(d^G_{\mathit{eth}}\right)$, where $P_i\left(d^G_{\mathit{eth}}\right)$ denotes the $i^{th}$ percentile of $d^G_{\mathit{eth}}$. In our study, we repeated this process using $i=10, 20, 30, 40, 50$, but in this figure we only illustrate the case where $i=50$.
 }
\label{fig:CEM_infographic}
\end{figure}

\begin{figure}[H]%
\centering
\includegraphics[width=.72\textwidth]{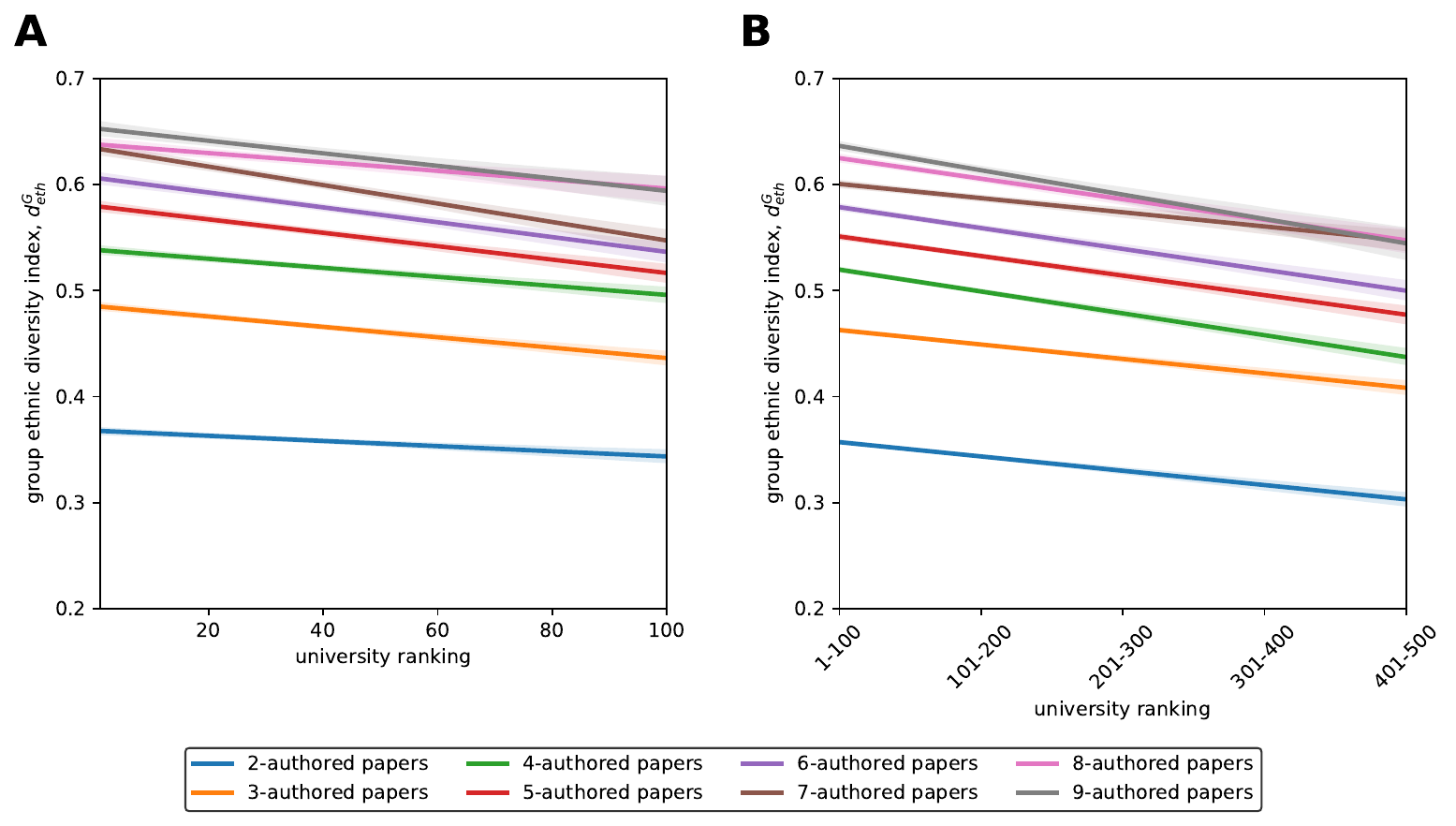}
 \caption{ {University ranking against group ethnic diversity.} \textbf{(A)} $d^G_{\mathit{eth}}$ for universities ranked $1, 2, \ldots, 100$. \textbf{(B)} $d^G_{\mathit{eth}}$ for universities whose ranking falls in one of 5 bins: 1-100; 101-200; 201-300; 301-400; 401-500. In both subfigures, we control for the number of authors per paper. A significant negative correlation between $d^G_{\mathit{eth}}$ and university ranking is found in all cases ($p<0.001$).}
\label{fig:unirank_vs_groupdiv}
\end{figure}

\begin{figure}[H]%
\centering
\includegraphics[width=.72\textwidth]{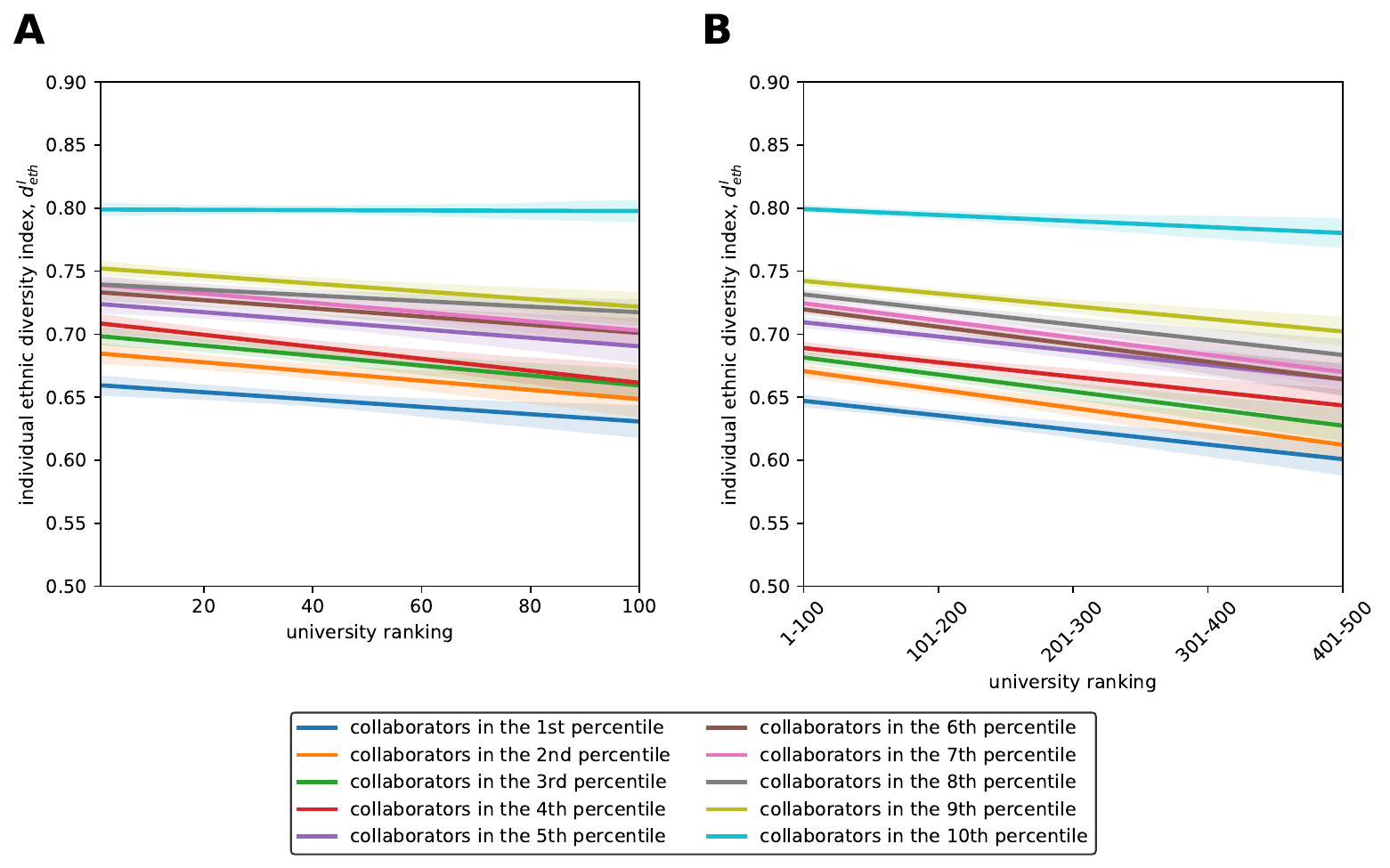}
 \caption{ {University ranking against individual ethnic diversity.} \textbf{(A)} $d^I_{\mathit{eth}}$ for universities ranked $1, 2, \ldots, 100$. \textbf{(B)} $d^I_{\mathit{eth}}$ for universities whose ranking falls in one of 5 bins: 1-100; 101-200; 201-300; 301-400; 401-500. In both subfigures, we control for the number of collaborators per scientist. A significant negative correlation between $d^I_{\mathit{eth}}$ and university ranking is found in all cases ($p<0.001$).}
\label{fig:unirank_vs_inddiv}
\end{figure}


\section*{Supplementary Tables}

\begin{table}[H]
\caption{Summary of main notation}\label{tab:notation}
\begin{center}
{\fontsize{10}{10}\selectfont{
\begin{tabular}{| p{2.2cm}|p{11.3cm}|}
\hline
{\fontsize{12}{12}\selectfont{\bf Notation}} & {\fontsize{12}{12}\selectfont{\bf Description}} \Tstrut\\[0.14cm]
\hline
$S$ & The set of scientists under consideration\Tstrut\\[0.14cm]
$P$ & The set of papers under consideration\\[0.14cm]
$\mathit{Papers}(s_i)$ & The set of papers of scientist $s_i$ \\[0.14cm]
$\mathit{Authors}(p_j)$ & The set of authors of paper $p_j$ \\[0.14cm]
$\mathit{Keywords}(p_j)$ & The set of keywords in paper $p_j$ \\[0.14cm]
$\mathit{Gini}(M)$ & \textit{Gini impurity} of multiset $M$; see Equation~\eqref{eqn:GiniImpurity}\\[0.14cm]
$\mathit{Disciplines}$ & The set of 19 scientific disciplines in \textit{Microsoft Academic Graph (MAG)} \\[0.14cm]
$\mathit{dsp}(s_i)$ & Discipline of scientist $s_i$ according to \textit{Microsoft Academic Graph (MAG)}; see Equation~\eqref{eqn:Field}\\[0.14cm]
$\mathit{eth}(s_i)$ & Ethnicity of scientist $s_i$ according to the \textit{Name Ethnicity Classifier}\\[0.14cm]
$\mathit{gen}(s_i)$ & Gender of scientist $s_i$ according to \textit{Genderize.io}\\[0.14cm]
$\mathit{age}(s_i)$ & Academic age of scientist $s_i$, measured by subtracting the publication year of the first paper of $s_i$ from the year 2009\\[0.14cm]
$\mathit{aff}(s_i,p_j)$ & Affiliation of scientist $s_i$ in paper $p_j$ according to \textit{Microsoft Academic Graph}\\[0.14cm]
$d^G_{\mathit{dsp}}$ & Group discipline diversity index; see Equation~\eqref{eqn:GroupDiversityIndex}, where $x=\mathit{dsp}$ \\[0.14cm]
$d^G_{\mathit{eth}}$ & Group ethnic diversity index; see Equation~\eqref{eqn:GroupDiversityIndex}, where $x=\mathit{eth}$ \\[0.14cm]
$d^G_{\mathit{gen}}$ & Group gender diversity index; see Equation~\eqref{eqn:GroupDiversityIndex}, where $x=\mathit{gen}$ \\[0.14cm]
$d^G_{\mathit{age}}$ & Group age diversity index; see Equation~\eqref{eqn:GroupDiversityIndex}, where $x=\mathit{age}$ \\[0.14cm]
$d^G_{\mathit{aff}}$ & Group affiliation diversity index; see Equation~\eqref{eqn:GroupDiversityIndex}, where $x=\mathit{aff}$ \\[0.14cm]
$\langle d^G_x\rangle$ & Average $d^G_x$ taken over a set of papers, where $x \in \{\mathit{eth, gen, age, dsp, aff}\}$\\[0.14cm]
$d^I_{\mathit{dsp}}$ & Individual discipline diversity index; see Equation~\eqref{eqn:IndividualDiversityIndex:forScientist}, where $x=\mathit{dsp}$ \\[0.14cm]
$d^I_{\mathit{eth}}$ & Individual ethnic diversity index; see Equation~\eqref{eqn:IndividualDiversityIndex:forScientist}, where $x=\mathit{eth}$ \\[0.14cm]
$d^I_{\mathit{gen}}$ & Individual gender diversity index; see Equation~\eqref{eqn:IndividualDiversityIndex:forScientist}, where $x=\mathit{gen}$ \\[0.14cm]
$d^I_{\mathit{age}}$ & Individual age diversity index; see Equation~\eqref{eqn:IndividualDiversityIndex:forScientist}, where $x=\mathit{age}$ \\[0.14cm]
$d^I_{\mathit{aff}}$ & Individual affiliation diversity index; see Equation~\eqref{eqn:IndividualDiversityIndex:forScientist}, where $x=\mathit{aff}$ \\[0.14cm]
$\langle d^I_x\rangle$ & Average $d^I_x$ taken over a set of individuals, where $x \in \{\mathit{eth, gen, age, dsp, aff}\}$\\[0.14cm]
$d^I_x(p_j)$ & Average of $d^I_{\mathit{eth}}$ over all authors of $p_j$, where $x \in \{\mathit{eth, gen, age, dsp, aff}\}$; see Equation~\eqref{eqn:IndividualDiversityIndex:forPaper}\\[0.14cm]
$\big<d^I_{x}\big>_{\textnormal{paper}}$ & An alternative notation of $d^I_x(p_j)$ which is often used when the paper $p_j$ is clear from the context\\[0.14cm]
$c^G_5(p_j)$ & Number of citations that paper $p_j$ accumulates 5 years after publication \\[0.14cm]
$\langle c^G_5\rangle$ & Average $c^G_5$ taken over a set of papers \\[0.14cm]
$c^I_5(s_i)$ & Number of citations that scientist $s_i$ accumulates on average from his/her papers 5 years after publication; see Equation~\eqref{eq:individualImpact} \\[0.14cm]
$\langle c^I_5\rangle$ & Average $c^I_5$ taken over a set of individuals\\[0.14cm]
$P_i\left(d^G_{\mathit{eth}}\right)$ & The $i^{th}$ percentile of $d^G_{\mathit{eth}}$\\
[0.14cm]

\hline
\end{tabular}
}}
\end{center}
\end{table}


\begin{table}[H]
{\fontsize{10}{10}\selectfont{
\caption{Pearson's $r$ and $p$ values corresponding to each subfield in Supplementary Figure~\ref{fig:paperindices_subfields}.
}
\label{tab:GroupDiversityIndexCorr}
\begin{center}
\begin{tabular}{lrlrlrlrlrl}
\toprule
Field &  $r_{\mathit{eth}}$ &  $p_{\mathit{eth}}$ &  $r_{\mathit{age}}$ &  $p_{\mathit{age}}$ &  $r_{\mathit{dsp}}$ &  $p_{\mathit{dsp}}$ &  $r_{\mathit{aff}}$ &  $p_{\mathit{aff}}$ &  $r_{\mathit{gen}}$ &  $p_{\mathit{gen}}$\\
\midrule
         Sustainable Energy &   0.06 &       0.00 &   0.06 &       0.00 &   0.09 &       0.00 &  -0.05 &       0.00 & -0.01 &       0.14$^{\dagger}$\\
       Mechanical Engineering &   0.06 &       0.00 &   0.04 &       0.00 &   0.04 &       0.00 &  -0.06 &       0.00 &  -0.02 &       0.04 \\
               Bioinformatics &   0.01 &       0.56$^{\dagger}$ &   0.02 &       0.00 &   0.02 &       0.00 &   0.00 &       0.68$^{\dagger}$ &  -0.03 &       0.00 \\
                   Cardiology &   0.05 &       0.00 &   0.08 &       0.00 &   0.04 &       0.00 &   0.02 &       0.01  &  -0.01 &       0.00 \\
                   Psychiatry &   0.03 &       0.00 &   0.24 &       0.00 &   0.05 &       0.00 &  -0.02 &       0.12$^{\dagger}$  &   0.05 &       0.00 \\
                      Nursing &   0.04 &       0.00 &   0.02 &       0.00 &   0.04 &       0.00 &  -0.02 &       0.12$^{\dagger}$ &  -0.00 &       0.93$^{\dagger}$ \\
                   Accounting &   0.07 &       0.00 &   0.03 &       0.02 &  -0.02 &       0.26$^{\dagger}$ &  -0.02 &       0.44$^{\dagger}$ &  -0.01 &       0.66$^{\dagger}$ \\
                    Marketing &   0.05 &       0.03 &   0.02 &       0.00 &   0.02 &       0.06$^{\dagger}$ &   0.02 &       0.51$^{\dagger}$  &  -0.00 &       0.92$^{\dagger}$ \\
   Educational Administration &   0.20 &       0.00 &   0.07 &       0.00 &   0.07 &       0.00 &   0.06 &       0.04 &   0.04 &       0.05 \\
       Language \& Linguistics &   0.07 &       0.00 &   0.07 &       0.00 &   0.03 &       0.02 &   0.02 &       0.47$^{\dagger}$  &   0.06 &       0.00 \\
   Philosophy &   0.08 &       0.03 &   0.07 &       0.00 &   0.03 &       0.03 &  -0.04 &       0.21$^{\dagger}$ &   0.00 &       0.84$^{\dagger}$ \\
   Drama &   0.00 &       0.35$^{\dagger}$ &   0.22 &       0.00 &  -0.14 &       0.00 &   0.01 &       0.90$^{\dagger}$ &   0.04 &       0.51$^{\dagger}$ \\
Mathematical Optimization &   0.07 &       0.00 &   0.02 &       0.02 &   0.03 &       0.00 &   0.00 &       0.73$^{\dagger}$  &  -0.00 &       0.80$^{\dagger}$ \\
              Fluid Mechanics &   0.02 &       0.02 &  -0.01 &       0.18$^{\dagger}$ &   0.02 &       0.00 &  -0.02 &       0.01 &  -0.04 &       0.00 \\
         Mathematical Physics &   0.03 &       0.00 &   0.02 &       0.00 &   0.01 &       0.15$^{\dagger}$ &   0.04 &       0.00  &  -0.04 &       0.00 \\
        Political Science &   0.05 &       0.00 &   0.08 &       0.00 &   0.05 &       0.00 &  -0.04 &       0.03  &  -0.01 &       0.47$^{\dagger}$ \\
        Sociology &   0.03 &       0.22$^{\dagger}$ &   0.04 &       0.00 &  -0.01 &       0.23$^{\dagger}$ &  -0.02 &       0.34$^{\dagger}$  &   0.04 &       0.00 \\
       History &   0.20 &       0.00 &   0.04 &       0.00 &   0.05 &       0.00 &  -0.04 &       0.17$^{\dagger}$  &   0.01 &       0.14$^{\dagger}$ \\
 Oil, Petroleum \& Nat. Gas &   0.15 &       0.00 &   0.05 &       0.00 &  -0.07 &       0.00 &   0.14 &       0.00 &  -0.02 &       0.30$^{\dagger}$ \\
 Nanotechnology &   0.11 &       0.00 &   0.05 &       0.00 &   0.06 &       0.00 &  -0.02 &       0.12$^{\dagger}$&  -0.01 &       0.22$^{\dagger}$ \\
        Materials Engineering &   0.02 &       0.00 &   0.02 &       0.00 &   0.05 &       0.00 &   0.00 &       0.76$^{\dagger}$  &  -0.01 &       0.04 \\
     Geology &   0.04 &       0.00 &   0.00 &       0.65$^{\dagger}$ &   0.06 &       0.00 &   0.02 &       0.14$^{\dagger}$ &  -0.03 &       0.00 \\
       Environmental Sciences &   0.05 &       0.00 &   0.04 &       0.00 &   0.05 &       0.00 &   0.01 &       0.22$^{\dagger}$  &  -0.01 &       0.00 \\ 
                      Ecology &   0.01 &       0.29$^{\dagger}$ &   0.06 &       0.00 &   0.02 &       0.00 &  -0.00 &       0.84$^{\dagger}$ &  -0.02 &       0.00 \\
\bottomrule
$^{\dagger}$ $p \geq 0.05$\\
\end{tabular}
\end{center}
}}
\end{table}

\newpage

\begin{table}
\renewcommand{\arraystretch}{1.5}
\caption{{{MAG subset definitions}. The total number of papers considered is: $9,472,439$ (those include the $1,045,401$ papers in the dataset $\mathcal{D}$). In contrast, the total number of scientists considered is around 6 million: $(1,529,279) + (5,103,877) - (\textnormal{overlap between the two sets})$.}}\label{tab:summaryOfDatasets}
\begin{tabular}{|C{5cm}|C{6cm}|C{4cm}|}
\hline
\vspace*{-0.5cm}{\textbf{Dataset}}
&
\vspace*{-0.5cm}{\textbf{Filter}}
&
\vspace*{-0.5cm}{\textbf{Set Size}}
\\ \hline\hline
\vspace*{-0.5cm}{Main dataset ($\mathcal{D}$). This is the dataset that will be used by default in all analyses unless stated otherwise.}
&
\vspace*{-0.5cm}{Of all the papers in the Microsoft Academic Graph (MAG) dataset, we considered all papers from the top five journals from 3 randomly selected subfields from each of the 8 main fields. {Then, we removed single-authored and review papers, controlled for English speaking countries, and only retained papers published between 1958 and 2009.}}
&
\vspace*{-0.5cm}{1,045,401 papers, authored by 1,529,279 unique authors.}
\\ \hline
\vspace*{-0.5cm}{
Dataset to measure group gender diversity index, $d^G_{\mathit{gen}}$}
&
\vspace*{-0.5cm}{
Papers in $\mathcal{D}$ where the gender of all authors are known with 90\% certainty (using Genderize.io).
}
&
\vspace*{-0.5cm}{
460,238 papers}
\\ \hline
\vspace*{-0.5cm}{
Dataset to measure group discipline diversity index, $d^G_{\mathit{dsp}}$}
&
\vspace*{-0.5cm}{
Papers in $\mathcal{D}$ where the discipline of all authors in a paper is clear and known (i.e. not interdisciplinary) was used.}
&
\vspace*{-0.5cm}{
568,269 papers}
\\ \hline
\vspace*{-0.5cm}{
Dataset to measure group affiliation diversity index, $d^G_{\mathit{aff}}$}
&
\vspace*{-0.5cm}{
Papers in $\mathcal{D}$ where each author has exactly one affiliation}
&
\vspace*{-0.5cm}{
207,899 papers
}
\\ \hline
\vspace*{-0.5cm}{
Dataset to measure the individual diversity indices, $d^I_x$}
&
\vspace*{-0.5cm}{
Scientists in $\mathcal{D}$ with at least 10 collaborators}
&
\vspace*{-0.5cm}{
766,338 scientists  with a total of 5,103,877 collaborators taken from 9,472,439 papers}
\\ \hline
\end{tabular}
\end{table}

\clearpage
\newpage

\begin{table}[H]
{\fontsize{10}{10}\selectfont{
\caption{Pearson's $r$ and $p$ values corresponding to each subfield in Supplementary Figure~\ref{fig:authorindices_subfields}.
}
\label{tab:IndividualDiversityIndexCorr}
\begin{center}
\begin{tabular}{lrlrlrlrlrl}
\toprule
Field &  $r_ {\mathit{eth}}$ &  $p_{\mathit{eth}}$ & $r_{\mathit{age}}$ &  $p_{\mathit{age}}$ & $r_{\mathit{dsp}}$ &  $p_{\mathit{dsp}}$ & $r_{\mathit{aff}}$ &  $p_{\mathit{aff}}$ & $r_{\mathit{gen}}$ &  $p_{\mathit{gen}}$ \\
\midrule
          Sustainable Energy &             0.05 &           0.00 &             0.04 &           0.00 &            -0.02 &           0.55$^{\dagger}$ &            -0.03 &           0.51$^{\dagger}$ &             0.00 &           0.50$^{\dagger}$ \\
       Mechanical Engineering &             0.06 &           0.00 &             0.07 &           0.00 &             0.01 &           0.79$^{\dagger}$ &            -0.06 &           0.15$^{\dagger}$ & 0.00 & 0.39$^{\dagger}$\\
               Bioinformatics &             0.02 &           0.00 &             0.01 &           0.00 &            -0.02 &           0.23$^{\dagger}$ &            -0.16 &           0.00& -0.00 & 0.01 \\
                   Cardiology &             0.06 &           0.00 &             0.05 &           0.00 &             0.05 &           0.12$^{\dagger}$ &             0.07 &           0.00& 0.00 & 0.75$^{\dagger}$ \\
                   Psychiatry &             0.03 &           0.00 &             0.04 &           0.00 &             0.05 &           0.37$^{\dagger}$ &             0.10 &           0.08$^{\dagger}$ & 0.02 & 0.00 \\
                      Nursing &             0.03 &           0.00 &             0.04 &           0.00 &             0.06 &           0.32$^{\dagger}$ &            -0.34 &           0.00 & 0.02 & 0.00\\
                   Accounting &             0.06 &           0.00 &             0.04 &           0.00 &            -0.01 &           0.86$^{\dagger}$ &             0.03 &           0.61$^{\dagger}$& -0.02 &           0.11$^{\dagger}$ \\
                    Marketing &             0.01 &           0.05 &             0.04 &           0.00 &             0.12 &           0.13$^{\dagger}$ &             0.10 &           0.28$^{\dagger}$ & 0.00 &           0.81$^{\dagger}$  \\
   Educational Administration &             0.10 &           0.00 &             0.06 &           0.00 &             0.23 &           0.00 &            -0.29 &           0.00 & 0.05 &           0.00 \\
       Language \& Linguistics &             0.07 &           0.00 &             0.08 &           0.00 &             0.14 &           0.20$^{\dagger}$&             0.10 &           0.34$^{\dagger}$ & 0.03 &       0.01 \\
                   Philosophy &             0.06 &           0.00 &             0.03 &           0.08$^{\dagger}$ &             0.27 &           0.04 &            -0.34 &           0.00 & 0.01 &           0.75$^{\dagger}$ \\
                        Drama &             0.13 &           0.00 &             0.11 &           0.05$^{\dagger}$ &             0.07 &           0.70$^{\dagger}$&             0.02 &           0.91$^{\dagger}$ & 0.02 &           0.79$^{\dagger}$ \\
    Mathematical Optimization &             0.06 &           0.00 &             0.05 &           0.00 &             0.07 &           0.20$^{\dagger}$&            -0.09 &           0.07$^{\dagger}$ & -0.00 &   0.62$^{\dagger}$ \\
              Fluid Mechanics &             0.06 &           0.00 &             0.04 &           0.00 &             0.07 &           0.26$^{\dagger}$&            -0.04 &           0.53$^{\dagger}$ & -0.01 &           0.00\\
         Mathematical Physics &             0.04 &           0.00 &             0.01 &           0.00 &             0.04 &           0.42$^{\dagger}$&             0.07 &           0.17$^{\dagger}$ & -0.01 &           0.00\\
            Political Science &             0.08 &           0.00 &             0.06 &           0.00 &             0.25 &           0.02 &            -0.14 &           0.21$^{\dagger}$ & -0.01 &           0.65$^{\dagger}$\\
                    Sociology &             0.06 &           0.00 &             0.04 &           0.00 &             0.05 &           0.67$^{\dagger}$ &             0.09 &           0.42$^{\dagger}$ & 0.03 &           0.02 \\
                      History &             0.04 &           0.00 &             0.07 &           0.00 &            -0.03 &           0.88$^{\dagger}$ &            -0.40 &           0.29$^{\dagger}$ & 0.01 &           0.25$^{\dagger}$\\
 Oil, Petroleum \& Nat. Gas &             0.14 &           0.00 &             0.10 &           0.00 &             0.02 &           0.78$^{\dagger}$ &             0.16 &           0.02 & -0.03 &           0.00 \\
               Nanotechnology &             0.06 &           0.00 &             0.04 &           0.00 &             0.09 &           0.13$^{\dagger}$&             0.01 &           0.89$^{\dagger}$ & 0.01 &           0.00\\
        Materials Engineering &             0.02 &           0.00 &             0.02 &           0.00 &            -0.05 &           0.32$^{\dagger}$&             0.05 &           0.29$^{\dagger}$ & 0.01 &           0.00\\
                      Geology &             0.03 &           0.00 &             0.06 &           0.00 &             0.03 &           0.57$^{\dagger}$&             0.09 &           0.15$^{\dagger}$ & 0.03 &           0.00\\
       Environmental Sciences &             0.08 &           0.00 &             0.07 &           0.00 &            -0.04 &           0.51$^{\dagger}$ &             0.09 &           0.19$^{\dagger}$ &             -0.01 &           0.00 \\
                      Ecology &             0.04 &           0.00 &             0.05 &           0.00 &             0.11 &           0.05$^{\dagger}$ &             0.07 &           0.19$^{\dagger}$&             -0.00 &           0.90$^{\dagger}$ \\
\bottomrule
$^{\dagger}$ $p \geq 0.05$\\
\end{tabular}
\end{center}
}}
\end{table}


\begin{table}[H]
{\fontsize{10}{10}\selectfont{
\caption{{Results of coarsened exact matching on group ethnic diversity}. $T$ and $C$ are the treatment and control populations respectively; $T'$ and $C'$ are the populations of matched treatment and matched control papers respectively; $\mathcal{L}_1$ is  the multivariate imbalance statistic~\cite{iacus2012causal}; $\delta$ is the relative impact gain of $T'$ over $C'$, i.e., $\delta = 100\times(\langle c^G_5 \rangle_{T'} - \langle c^G_5 \rangle_{C'})/\langle c^G_5 \rangle_{C'}$. A t-test shows that $\delta$ is statistically significant; see the resulting $p$-values. Since the academic impact $\langle c^G_5 \rangle$ is sensitive to extremal values, we bootstrap a 95\% confidence interval ($CI_{.95}$). Note that the confounding factor ``university ranking'' corresponds to the \emph{highest ranked} of all universities in the paper, as opposed to the \emph{average rank} for all universities in the paper, as is the case in Table~2. For more details, see Supplementary Note~\ref{sec:CEM}.}
\label{tab:CEMGroup}
\begin{center}
\begin{tabular}{lcccccccccccc}
\toprule
 & $|T|$ & $|C|$ &  $|T'|$ & $|C'|$ &  $\mathcal{L}_1$ & $\delta$ & $CI_{.95}$ &  $p$\\
\midrule
$T : d^G_{eth} > P_{90}(d^G_{eth})$ & 17,802 & 45,710 & 14,876 & 16,180 & 0.38 & 11.33 & [9.12, 13.45]&  0.002\\
$C : d^G_{eth} \leq P_{10}(d^G_{eth})$\\
\midrule
$T: d^G_{eth} > P_{80}(d^G_{eth})$  & 24,827 & 45,710  & 20,808 & 16,294 & 0.38 & 11.56  & [9.52, 13.39]&  0.0001\\
$C: d^G_{eth} \leq P_{20}(d^G_{eth})$\\
\midrule
$T: d^G_{eth} > P_{70}(d^G_{eth})$ & 56,662 & 58,889 & 53,588 & 39,376 & 0.25 & 5.58 & [4.23, 6.75]& 0.0066\\
$C: d^G_{eth} \leq P_{30}(d^G_{eth})$\\
\midrule

$T: d^G_{eth} > P_{60}(d^G_{eth})$  & 63,129 & 78,340 & 58,903  & 58,411 & 0.28 & 6.50 & [5.48, 7.47]& 0.0002\\
$C: d^G_{eth} \leq P_{40}(d^G_{eth})$\\
\midrule
$T: d^G_{eth} > P_{50}(d^G_{eth})$ & 63,129 & 127,629 & 59,474 & 70,958 & 0.26 & 3.86 & [2.96, 4.21 ]& 0.042\\
$C: d^G_{eth} \leq P_{50}(d^G_{eth})$\\
\midrule
\end{tabular}
\end{center}
}}
\end{table}

\newpage



{\noindent\Large \textbf{Supplementary Notes}}

\section{\hspace{-.5cm}.\hspace{.35cm}The Data}\label{SM:data}

\subsection*{The Collaboration Network}
The data used for this study was obtained on October 2015 from the Microsoft Academic Graph (MAG) database.\footnote{\tt https://www.microsoft.com/en-us/research/project/microsoft-academic-graph/} This is a dataset consisting of scientific publications, their citation records, date of publication, information regarding the authorship (such as name and affiliation), publication venue and more. 
The dataset also contains a citation network in which every node represents a paper and every directed link represents a citation. While the number of citations of any given paper is not provided explicitly by the dataset, it can easily be calculated from the citation network. More important, the dataset specifies the keywords in each paper, as well as the position of each such keyword in a field-of-study hierarchy, the highest level of which is comprised of 19 disciplines.\footnote{Note that some keywords fall under multiple disciplines. For instance, according to the dataset, the keyword ``Fast fission'' has a 50\% match with Physics and a 50\% match with Chemistry.}

Unfortunately, the dataset suffers from three limitations: (i) it does not specify the publication venue's field of science; (ii) it does not specify the \textit{ethnicity} of each scientist; and (iii) it does not specify the \textit{gender} of each scientist. In the following three Supplementary Notes, we show how to overcome these limitations.

\subsection*{Acquiring the Field of Science of Each Publication Venue}
\label{SM:GoogleScholar}
To address limitation (i) of Microsoft Academic Graph, we refer to Google Scholar Metrics.\footnote{\tt https://scholar.google.com/citations?view\_op=top\_venues} Here, journals are categorized into 8 main fields of science, and each such field is divided into multiple subfields. A list of the top 20 publication venues are listed for each subfield. We considered five top journals from 3 randomly selected subfields from each of the 8 main fields. The main fields and their subfields are as follows:

\begin{multicols}{2}
\fontsize{12}{12}\selectfont{
\begin{enumerate}
\item \textbf{Engineering \& Computer Science}\itemsep -0.5em\vspace{-0.5em}
	\begin{enumerate}
    \item Mechanical Engineering
    \item Sustainable Energy
    \item Bioinformatics
    \end{enumerate}

\item \textbf{Health \& Medical Sciences}\itemsep -0.5em\vspace{-0.5em}
	\begin{enumerate}
    \item Cardiology
    \item Psychiatry
    \item Nursing
    \end{enumerate}
    
\item \textbf{Business, Economics \& Management}\itemsep -0.5em\vspace{-0.5em}
	\begin{enumerate}
    \item Accounting
    \item Marketing
    \item Educational Administration
    \end{enumerate}
    
\item \textbf{Humanities, Literature \& Arts}\itemsep -0.5em\vspace{-0.5em}
	\begin{enumerate}
    \item Language \& Linguistics
    \item Drama
    \item Philosophy
    \end{enumerate}
    
\item \textbf{Physics \& Mathematics}\itemsep -0.5em\vspace{-0.5em}
	\begin{enumerate}
    \item Mathematical Optimization
    \item Mathematical Physics
    \item Fluid Mechanics
    \end{enumerate}

\item \textbf{Social Sciences}\itemsep -0.5em\vspace{-0.5em}
	\begin{enumerate}
    \item Political Science
    \item Sociology
    \item History
    \end{enumerate}
    
\item \textbf{Chemical \& Material Sciences}\itemsep -0.5em\vspace{-0.5em}
	\begin{enumerate}
    \item Oil, Petroleum \& Natural Gas
    \item Nanotechnology
    \item Material Engineering
    \end{enumerate}
    
\item \textbf{Life Sciences \& Earth Sciences}\itemsep -0.5em\vspace{-0.5em}
	\begin{enumerate}
    \item Environmental Sciences
    \item Geology
    \item Ecology
    \end{enumerate}
\end{enumerate}
}
\end{multicols}

For each journal, we  extracted the data on all papers published therein, and applied the following five filtering steps: (i) removed all single-authored papers (because we are interested in studying collaborations), (ii) controlled for English speaking countries (explained further in section Controlling for Countries), (iii) removed review papers (explained further in section Excluding Review Papers), (iv) extracted data only up to 2009 (the reason behind this is explained in Supplementary Note~\ref{SM:CitationsCount}), and (v) only retained papers published in or after 1958 (to deal with missing data issues; in all preceding years, there were multiple fields with no publications at all).

This yielded our main dataset, consisting of 1,045,401 papers, authored by 1,529,279 unique authors. This dataset will be filtered whenever necessary, as will be specified in the following sections (Supplementary Table~\ref{tab:summaryOfDatasets} summarizes all the filters used in our study).

Lastly, to avoid any potential confusion between a scientist's area of science, and a paper's area of science, we will use the term ``discipline'' when referring to the former, and use the term ``field'' when referring to the latter.

\subsection*{Classifying the Ethnicity of Each Scientist}
\label{SM:NameEthnicityClassifier}
To address limitation (ii) of Microsoft Academic Graph, we used the \textit{Name Ethnicity  Classifier}\footnote{\tt http://www.textmap.com/ethnicity/}\cite{ambekar2009name,ye2017nationality} to identify the ethnicity of each scientist. In particular, this classifier uses various machine-learning techniques to classify any given name into the following 13 ethnic groups (any unresolved names are marked as ``unknown''):
  \begin{enumerate}
  \itemsep-0.5em
  \item Asian, Greater East Asian, East Asian (or ``\textit{East Asian}'' for short);
  \item Asian, Greater East Asian, Japanese (or ``\textit{Japanese}'' for short);
  \item Asian, Indian Sub-Continent (or ``\textit{Indian Sub-Continent}'' for short);
  \item Greater African, Africans (or ``\textit{Africans}'' for short);
  \item Greater African, Muslim (or ``\textit{Muslim}'' for short);
  \item Greater European, British (or ``\textit{British}'' for short);
  \item Greater European, East European (or ``\textit{East European}'' for short);
  \item Greater European, Jewish (or ``\textit{Jewish}'' for short);
  \item Greater European, West European, French (or ``\textit{French}'' for short);
  \item Greater European, West European, Germanic (or ``\textit{Germanic}'' for short);
  \item Greater European, West European, Hispanic (or ``\textit{Hispanic}'' for short);
  \item Greater European, West European, Italian (or ``\textit{Italian}'' for short);
  \item Greater European, West European, Nordic (or ``\textit{Nordic}'' for short).
  \end{enumerate}

As can be seen, the term ``ethnicity" is used here in its broader sense, where an ethnic group is defined as \emph{``a social group that shares a common and distinctive culture, religion, language, or the like"}.\footnote{http://www.dictionary.com/browse/ethnicity?s=t} The Name Ethnicity Classifier \cite{ambekar2009name,ye2017nationality, wired} has an overall accuracy\footnote{While the authors report the results of each ethnicity independently, the overall accuracy can easily be computed from these results.} of 80\%, which is quite impressive given that the classifier depends solely on the individual's name. Importantly, this accuracy is measured over 20 million distinct names, comprising over 100 million individual entity references \cite{ambekar2009name}. Admittedly, an accuracy of 80\% means that some names will be misclassified. Nevertheless, unlike conventional methods where ethnicity is identified manually, this classifier allows for conducting studies at an unprecedented scale, e.g., involving millions of names. In our case, we were able to obtain the ethnicity of every single scientist in our study.

\subsection*{Identifying the Gender of Each Scientist}
\label{SM:GenderIdentification}

To address limitation (iii) of Microsoft Academic Graph, we needed to identify the gender of each scientist in our dataset. To this end, a number of alternative methods have been proposed in the literature to identify the gender (either male or female) of any given individual based solely on his/her first name \cite{wais2016gender,west2013role,lariviere2013bibliometrics}. Out of all those alternatives, a software tool called ```\emph{Genderize.io}'', which is available at: {\tt https://genderize.io/}, was shown to be the most reliable \cite{wais2016gender}. Note that gender identification based solely on first name is indeed very challenging (if not impossible, in some cases), due mainly to the fact that some names are unisex. As such, it is perhaps not surprising that 47.71\% of the names tested on Genderize.io were unclassified, i.e., the tool returned ``unknown'' for each such name \cite{wais2016gender}. Nevertheless, this tool outperformed the alternative methods, for which the number of unclassified names exceeded 84\%. As for the names that were classified, the ``error rate'' represents the percentage of names whose classification was incorrect. While the  alternative methods had an error rate greater that 32\%, Genderize.io had an error rate of just 7\%, which is impressive given that the input to this gender-identification tool is merely the first name of the individual in question. Admittedly, an error rate of 7\% means that some genders will be misclassified. Nevertheless, unlike conventional methods where gender is identified manually, this automated tool allows for conducting studies at an unprecedented scale, covering thousands, or even millions of names. Using Genderize.io, we were able to classify the gender of 3,183,911 scientists. To further increase our confidence of the gender classification, we considered only the 2,046,359 names that were classified with at least 90\% confidence.

\subsection*{Controlling for Countries}
\label{SM:ControlForCountries}

To control for countries, we consider papers of which the majority of the authors' affiliations belong to any of the following countries: USA, UK, Canada and Australia. The rationale behind this choice is threefold:

\begin{enumerate}
\item These are predominantly English-speaking countries.\footnote{Although Canada has two official languages, namely English and French, 56.9\% of the Canadian population report English as their mother tongue; see: www.statcan.gc.ca.} As English is widely considered the universal language of science,\footnote{\tt \fontsize{7}{7}\selectfont{https://www.researchtrends.com/issue-31-november-2012/the-language-of-future-scientific-communication/}} limiting our study to English-speaking countries should cover a wide range of cultural, ethnic and national backgrounds.
\item These countries have ethnically diverse populations and higher-education systems that attract large numbers of international students and faculty members. In contrast, universities in many other countries (such as Japan and China) have student populations of which the vast majority are of a single ethnic group.
\item Universities from those four countries form a significant proportion of the world's leading institutions (as reported in the Times Higher Education 2018 rankings, 95\% of the world's top 20, and 63\% of the top 100 universities are from these four countries\footnote{\tt \fontsize{7}{7}\selectfont{https://www.timeshighereducation.com/world-university-rankings/2018/world-ranking\#!/}}).
\end{enumerate}

To put it differently, most papers (co)authored in these four countries are arguably (i) written in the same language, (ii) produced in environments that permit the formation of diverse teams and (iii) relatively more likely to produce high-impact research.
The above factors help to ensure that the papers studied are, in general, highly impactful and  comparable. Of all the affiliations present in the Microsoft Academic Graph dataset, 5,899 (out of a total of 19,788) were manually verified to be based in one of the aforementioned four countries. 

\subsection*{Excluding Review Papers}
\label{SM:ExcludingReviewPapers}
Review papers exhibit different statistics \cite{radicchi2008universality,radicchi2012testing,stringer2010statistical}, and could bias our results. As such, we excluded from our analysis any review papers that we could find  based on tell-tale words that could be found in the keywords of the paper, such as ``literature review'', ``literature'', or ``survey''. Following this process, 11,367 review papers were found and removed from our dataset.

\newpage

\section{\hspace{-.5cm}.\hspace{.35cm}Quantifying Diversity}\label{SM:QuantifyingDiversity}

This note starts by discussing the five classes of diversity that are considered in our study, followed by a discussion of the group and individual diversity indices, respectively. A summary of all notation is provided in Supplementary Table~\ref{tab:notation}.

\subsection*{Classes of Diversity}\label{SM:TypesofDiversity}
When exploring diversity in research collaborations, we investigate five classes of diversity:

\begin{enumerate}
\item \textbf{Ethnic diversity:} This class of diversity takes into consideration the ethnic background of each scientist. As described in Supplementary Note~\ref{SM:NameEthnicityClassifier}, we use the \textit{Name Ethnicity Classifier} to identify the ethnicity of each scientist.

\item \textbf{Gender diversity:} This class of diversity takes into consideration the gender of each scientist, which is identified using \textit{Genderize.io}. When studying gender diversity, we only include a paper if the gender of each of its author is identified by \textit{Genderize.io}; see Supplementary Note~\ref{SM:GenderIdentification} for more details on how gender is identified, and Supplementary Table~\ref{tab:summaryOfDatasets} for a summary of all datasets used in our study.

\item \textbf{Age diversity:} Here, ``age'' refers to the academic age of a scientist, which is measured  in each paper, $p$, by subtracting the year of the scientist's first paper from the year in which $p$ was published. The resulting dataset is then divided into the following bins:

\begin{itemize}
\itemsep-0.5em
\item Academic age group 0 : 0-9 years of experience;
\item Academic age group 1 : 10-19 years of experience; 
\item Academic age group 2 : 20-29 years of experience;
\item Academic age group 3 : 30-39 years of experience;
\item Academic age group 4 : 40-49 years of experience;
\item Academic age group 5 : $\geqslant$50 years of experience. 
\end{itemize}

\item \textbf{Discipline Diversity:} This class of diversity takes into account the co-authors' area of expertise. We determine the discipline of each scientist based on the keywords that are specified in his/her papers. This is made possible by the fact that the MAG dataset specifies the probability of each keyword belonging to any of the following 19 disciplines:

\begin{multicols}{2}
\begin{enumerate}
\itemsep-0.5em
\item[(1)] Art
\item[(2)] Biology 
\item[(3)] Business 
\item[(4)] Computer Science 
\item[(5)] Chemistry 
\item[(6)] Economics 
\item[(7)] Engineering 
\item[(8)] Environmental science
\item[(9)] Geography
\item[(10)] Geology 
\item[(11)] History
\item[(12)] Materials Science 
\item[(13)] Mathematics 
\item[(14)] Medicine	
\item[(15)] Philosophy 
\item[(16)] Physics 
\item[(17)] Political Science 
\item[(18)] Psychology 
\item[(19)] Sociology 
\end{enumerate}
\end{multicols}

Formally, the probability of scientist $s_i$ belonging to discipline $x_j$ is calculated as follows:
\begin{equation}\label{eqn:probabilityOfField}
P(\mathit{dsp}(s_i) = x_j) \ \ =\ \  \frac{
  \ \ \ \ \ \ \ \ \ \ \ \ \ \ \ \ \ 
  \sum\limits_{p\in \mathit{Papers}(s_i)}\ 
    \sum\limits_{w\in \mathit{Keywords}(p)} 
        P(\mathit{dsp}(w)=x_j)
  }{
  \sum\limits_{x_k\in\mathit{Disciplines}}\ 
    \sum\limits_{p\in \mathit{Papers}(s_i)}\ 
      \sum\limits_{w\in \mathit{Keywords}(p)}      
        P(\mathit{dsp}(w)=x_k)
  }
\end{equation}

\noindent where $\mathit{Papers}(s_i)$ denotes the set of papers of scientist $s_i$, $\mathit{Keywords}(p)$ denotes the set of keywords of paper $p$, $P(\mathit{dsp}(w)=x_j)$ denotes the probability that the keyword $w$ belongs to the discipline $x_j$, and $\mathit{Disciplines}$ denotes the set of the 19 disciplines in MAG. Then, the discipline of scientist $s_i$ is determined as follows:
\begin{equation}\label{eqn:Field}
\mathit{dsp}(s_i) = 
    \left\{
        \begin{array}{ll}
            \argmax\limits_{x_k \in \mathit{Disciplines}} P(\mathit{dsp}(s_i) = x_k) & \ \ \ \ \ \ \ \textnormal{if } \max\limits_{x_k \in \mathit{Disciplines}} P(\mathit{dsp}(s_i) = x_k) > 0.5\smallskip\smallskip\\
            \textnormal{``unknown''} & \ \ \ \ \ \ \ \textnormal{if } \max\limits_{x_k \in \mathit{Disciplines}} P(\mathit{dsp}(s_i) = x_k) \leq 0.5
        \end{array}
    \right.
\end{equation}

\noindent where $P(\mathit{dsp}(s_i) = x_k)$ is calculated as in Equation~\eqref{eqn:probabilityOfField}. We exclude from our analysis any paper of which the discipline of an author is ``unknown'' (see Supplementary Table~\ref{tab:summaryOfDatasets} for a summary of all the filters applied on our dataset).

\item \textbf{Affiliation Diversity:} This class of diversity takes into consideration the affiliations of the co-authors of a paper. Note that, where available, MAG specifies the affiliation of every author on every paper. As such, a scientist's affiliation may vary from one paper to another. We exclude any papers where an author has more than one affiliation or no affiliation at all. This way, having multiple affiliations on a paper indicates that it is the result of collaboration across different research entities (see Supplementary Table~\ref{tab:summaryOfDatasets}). 
\end{enumerate}

\subsection*{Measuring Diversity}

The diversity of any given group reflects the degree to which its members differ from one another. To study the relationship between this property and the success of the associated group, a numerical measure of group diversity is required. To this end, several metrics have been proposed, the majority of which fall into two main categories:
\begin{enumerate}
\item Metrics that measure diversity by quantifying the uncertainty in predicting the type of an element drawn randomly from the set in question. Such a metric is commonly known as the \emph{Shannon entropy} or the \emph{Shannon-Wiener Index}. Formally, given $k$ types, and a set $S$, the Shannon entropy is computed as in Equation~\eqref{eqn:shannon}, where $p_i(S)$ denotes the proportion of the elements of $S$ that are of the $i^{\text{th}}$ type.
\begin{equation}\label{eqn:shannon}
\mathit{Shannon}(S) = -\sum\limits_{i=1}^{k} p_i(S) \textnormal{ ln }p_i(S).
\end{equation}
\item Metrics that are designed to reflect the degree of concentration when the group members are classified into types \cite{simpson1949measurement}. Such a metric is commonly known as the \emph{Simpson index} in ecological literature, and as the \emph{Herfindahl-Hirschman index} in the economic literature \cite{herfindahl1950concentration}. It can also be found, with slight variations, in other fields under different names, including the \emph{probability of interspecific encounter} \cite{hurlbert1971nonconcept}, the \emph{Gini-Simpson index} \cite{jost2006entropy}, and the \emph{Gini impurity} \cite{rokach2005top}. The formula for the Gini impurity is computed as shown in Equation~\eqref{eqn:GiniImpurity}.
\end{enumerate}

For every paper in the entire MAG dataset, we measured the ethnic diversity in the group of authors using the Shannon entropy and using the Gini impurity. The two measures are plotted against each other in Supplementary Figure~\ref{fig:ShannonVsGiniImpurity}. As can be seen, the two are strongly correlated, with Pearson's $r=0.93$ and $p<0.0001$. Based on this, throughout the remainder of our study, we focus on just one of those measures, namely the Gini impurity, which will be explained in more detail in the following section.

\subsection*{Group Diversity Index}
\label{SM:GroupDiversityIndex}

In this section, we explain how the \textit{Gini impurity} \cite{bishop2006pattern} is used to measure the diversity in any given paper. To this end, we need to introduce some additional notation. Let $S$ and $P$ denote the set of scientists and the set of papers under consideration, respectively. Furthermore, let $\mathit{Authors}(p_j)\subseteq S$ denote the set of authors of paper $p_j$. Now, for any given scientist $s_i\in S$, let $\mathit{eth}(s_i)$, $\mathit{gen}(s_i)$, $\mathit{dsp}(s_i)$, and $\mathit{age}(s_i)$ denote the \textit{ethnicity}, the \textit{gender}, the \textit{discipline} and the \textit{academic age} of $s_i$, respectively. Similarly, let $\mathit{aff}(s_i,p_j)$ denote the \textit{affiliation} of scientist $s_i$ on paper $p_j$.\footnote{The affiliation of $s_i$ is denoted by $\mathit{aff}(s_i,p_j)$ rather than $\mathit{aff}(s_i)$ because the affiliation of a scientist may vary from one paper to another.} For details on how the ethnicity, gender, discipline, academic age, and affiliation are identified, section above on Classes of Diversity. Note that for any given paper, $p_j$, any set $\{x(s_i):s_i\in \mathit{Authors}(p_j)\}$ such that $x \in \{\mathit{eth, gen, age, dsp}\}$ is actually a multiset. Likewise, the set $\{\mathit{aff}(s_i,p_j):s_i\in \mathit{Authors}(p_j)\}$ is also a multiset. When dealing with multisets, we will use square brackets instead of curly ones. For instance, for any given paper, $p_j$, we could have: $[\mathit{eth}(s_i):s_i\in \mathit{Authors}(p_j)]=[\text{Japanese}, \text{British}, \text{British}]$, and have: $[\mathit{aff}(s_i,p_j):s_i\in \mathit{Authors}(p_j)]=[\text{Harvard}, \text{Harvard}, \text{Stanford}]$. 
For any given multiset, $M$, let $|M|$ denote the cardinality of $M$, let $\mathit{under}(M)$ denote the underlying set of $M$, and let $\mathit{multi}(m,M)$ denote the multiplicity of element $m$ in $M$. For example, given $M=[\text{Harvard}, \text{Harvard}, \text{Stanford}]$, we have: $|M|=3$, $\mathit{under}(M)=\{\text{Harvard}, \text{Stanford}\}$, $\mathit{multi}(\text{Harvard},M) = 2$ and $\mathit{multi}(\text{Stanford},M) = 1$. The \textit{Gini impurity} of a multiset, $M$, is then defined as:
\begin{equation}\label{eqn:GiniImpurity}
\mathit{Gini}(M) = 1 - \sum_{m\in \mathit{under}(M)} \mathit{proportion}(m,M)^2,
\end{equation}

\noindent where
$$
\mathit{proportion}(m, M) = \frac{\mathit{multi(m,M)}}{|M|}.
$$
With this notation in place, we are now ready to formally define our \textit{group diversity index}. In particular, for any given paper, $p_j\in P$, the group diversity index of $p_j$ is defined as follows, where the ``$G$'' in $d^G_x$ stands for ``\emph{Group}'':
\begin{equation}\label{eqn:GroupDiversityIndex}
d^G_x(p_j) = 
    \left\{
        \begin{array}{ll}
            \mathit{Gini}\left([x(s_i):s_i\in \mathit{Authors}(p_j)]\right) & \ \ \ \textnormal{if } x \in \{\mathit{eth}, \mathit{gen}, \mathit{dsp}, \mathit{age}\}\smallskip\smallskip\\
            \mathit{Gini}\left([x(s_i,p_j):s_i\in \mathit{Authors}(p_j)]\right) & \ \ \ \textnormal{if $x = \mathit{aff}$}
        \end{array}
    \right.
\end{equation}

\noindent We will often omit the paper, $p_j$, from the notation $d^G_x(p_j)$ and simply write  $d^G_x$ whenever the paper itself is clear from the context.

Next, we summarize our five group diversity indices, and specify the papers that were considered for each such index (out of all 1,045,401 papers published in our dataset):

\begin{enumerate}
\item $d^G_{\mathit{eth}}$---the ``\textit{group ethnic diversity index}''; we calculated this for all
papers in our dataset.
\item $d^G_{\mathit{gen}}$---the ``\textit{group gender diversity index}''; for any paper, we calculate this index only if the gender of each of author has been identified by Genderize.io.
\item $d^G_{\mathit{age}}$---the ``\textit{group age diversity index}''; this was calculated for all
papers in our dataset.
\item $d^G_{\mathit{dsp}}$---the ``\textit{group discipline diversity index}''; for this index, we exclude every paper of which an author's discipline is ``unknown'' according to Equation~\eqref{eqn:Field}.
\item $d^G_{\mathit{aff}}$---the ``\textit{group affiliation diversity index}'';  we calculated this index for every paper whose authors each have exactly one affiliation on the paper (i.e., we exclude papers of which an author has more than one affiliation, or no affiliation at all).
\end{enumerate}

\subsection*{Individual Diversity Index}
\label{SM:IndividualDiversityIndex}

For any given scientist, $s_i\in S$, the \textit{individual diversity index} of $s_i$ is defined as follows:
\begin{equation}\label{eqn:IndividualDiversityIndex:forScientist}
d^I_x(s_i) =     \left\{
        \begin{array}{ll}
            \mathit{Gini}\left(\underset{p_j\in \mathit{Papers}(s_i)}{\uplus} \big[x(s_k):s_k\in \mathit{Authors}(p_j) \setminus \{s_i\}\big]\right) & \ \ \ \textnormal{if } x \in \{\mathit{eth}, \mathit{gen}, \mathit{dsp}, \mathit{age}\}\smallskip\smallskip\\
            \mathit{Gini}\left(\underset{p_j\in\mathit{Papers}(s_i)}{\uplus} \big[x(s_k,p_j):s_j\in \mathit{Authors}(p_j) \setminus \{s_i\}\big]\right) & \ \ \ \textnormal{if $x = \mathit{aff}$}
        \end{array}
    \right.
\end{equation}

\noindent where ``$I$'' in $d^I_x$ stands for ``\emph{Individual}'', $\mathit{Papers}(s_i)$ denotes the set of papers of which scientist $s_i$ is an author, $\uplus$ denotes the multiset sum operation, and $\mathit{Gini}$ is defined as in Equation~\eqref{eqn:GiniImpurity}. We will clarify the notation through an example. Suppose that scientist $A$ is an author of just two papers, $p_1$ and $p_2$, such that:
\begin{itemize}
\item $\mathit{Authors}(p_1) = \{A,B,C\}$;
\item $\mathit{Authors}(p_2) = \{A,C,D\}$;
\item the ethnicities of B, C, and D are Japanese, British, and French, respectively.
\end{itemize}

\noindent Then we would have:
$$
{\fontsize{11}{11}\selectfont{
\begin{array}{lll}
\underset{p_j\in \mathit{Papers}(A)}{\uplus} \big[\mathit{eth}(s_k):s_k\in \mathit{Authors}(p_j) \setminus \{A\}\big]\ &=&\ [\mathit{eth}(B), \mathit{eth}(C)]\ \ \uplus\ \ [\mathit{eth}(C), \mathit{eth}(D)]\\
        &=&\ [\textnormal{Japanese}, \textnormal{British}]\ \uplus\ [\textnormal{British}, \textnormal{French}]\smallskip\\
        &=&\ [\textnormal{Japanese}, \textnormal{British}, \textnormal{British}, \textnormal{French}].
\end{array}
}}
$$

We will overload the notation by letting $d^I_x(p_i)$ denote the average individual diversity of the authors of paper $p_i$. More formally:
\begin{equation}\label{eqn:IndividualDiversityIndex:forPaper}
d^I_x(p_i) =\frac{ \sum\limits_{s_i\in\mathit{Authors}(p_i)} d^I_x(s_i) }{ \left|\mathit{Authors}(p_i)\right| },
\end{equation}

\noindent where $x \in \{\mathit{eth, gen, age, dsp, aff}\}$. To improve readability, we may write $\left<d^I_{\mathit{eth}}\right>_{\textnormal{paper}}$ instead of $d^I_x(p_i)$ when $p_i$ is clear from the context.  Moreover, when dealing with individual scientists, we will often write $d^I_x$ instead of $d^I_x(s_i)$ when $s_i$ is clear from the context.

To summarize, our five individual diversity indices are as follows:

\begin{enumerate}
\item $d^I_{\mathit{eth}}$---the ``\textit{individual ethnic diversity index}'';
\item $d^I_{\mathit{gen}}$---the ``\textit{individual gender diversity index}''; 
\item $d^I_{\mathit{age}}$---the ``\textit{individual age diversity index}'';
\item $d^I_{\mathit{dsp}}$---the ``\textit{individual discipline diversity index}'';
\item $d^I_{\mathit{aff}}$---the ``\textit{individual affiliation diversity index}''.
\end{enumerate}
 
Out of the 1,529,279 scientists in our dataset, we calculated the individual diversity index for those with at least ten collaborators each; this yielded a total of 766,338 scientists with 5,103,877 collaborators taken from 9,472,439 different papers. Furthermore, when studying the average individual diversity in each subfield, we excluded any scientist whose name appears in more than one subfield in our dataset. This led to the exclusion of 6.8\% of the scientists . For a summary of all filters applied on our dataset, see Supplementary Table~\ref{tab:summaryOfDatasets}.

\newpage

\section{\hspace{-.5cm}.\hspace{.35cm}The Randomized Baseline Model} \label{sec:randomizedBaselineModel}

\noindent In an attempt to isolate the effect of diversity from other confounding factors, we analyzed a randomized baseline model in which the scientists' ethnicities are shuffled while preserving all other characteristics. To explain how such a model is generated, we need some additional notation. For every paper, $p_j\in P$, let $\mathit{Field}(p_j)$ denote the subfield of $p_j$, and let $\mathit{Year}(p_j)$ denote the publication year of $p_j$. Furthermore, for every scientific subfield, $f$, number of authors, $n$, and publication year, $y$, let us denote by $S_{f,n,y}\subset S$ the set consisting of every author of a paper in subfield $f$ with $n$ authors, published in year $y$. More formally:
$$
S_{f,n,y}=\bigcup_{p_j\in P: \mathit{Field}(p_j)=f,|\mathit{Authors}(p_j)|=n,\mathit{Year}(p_j)=y}\{s\in \mathit{Authors}(p_j)\}.
$$
With this notation in place, we are ready to describe how the randomized baseline model is generated. To improve readability, our description will focus on ethnic diversity; for the other classes of diversity, simply replace $\mathit{eth}$ with the class of choice.

Recall that $\mathit{eth}(s)$ denotes the ethnicity of scientist $s$. Then, for each scientific subfield, $f$, number of authors, $n$, and publication year, $y$, the ethnicities in $S_{f,n,y}$ are shuffled as follows:
    \begin{enumerate}
    \item Create a list, $L_{f,n,y}$, such that $L_{f,n,y}[i] := \mathit{eth}(s_i), \forall s_i \in S_{f,n,y}$;
    \item Create a list, $L'_{f,n,y}$, which is a shuffled version of $L_{f,n,y}$;
    \item Set the ethnicities in the randomized model as follows: $\mathit{eth}(s_i) := L'_{f,n,y}[i], \forall s_i\in S_{f,n,y}$.  
    \end{enumerate}

\noindent The entire process was repeated 1,000 times, and the average \textit{group} ethnic diversity, $d^G_{\mathit{eth}}$, and average \textit{individual} ethnic diversity, $d^I_{\mathit{eth}}$, were used.

\newpage

\section{Scientific Impact: Citation Counts}
\label{SM:CitationsCount}
In their expansive study on scientific impact, Sinatra et al.~\cite{sinatra2016quantifying} studied the number of citations that a paper accumulates 10 years after publication, denoted by $c_{10}$; the same impact measure was later on used in \cite{Fortunatoeaao0185}. We follow a similar approach, but focus on 5 rather than 10 years. This way, we incorporate more recent papers in our study, which is particularly important since the majority of the papers in our study were published in recent years (Supplementary Figure~\ref{fig:PapersHist}). Based on this, as well as the fact that our dataset was obtained in October 2015, we only calculate $c_5$ for papers published between 1958 and 2009.

We distinguish between the number of citations that a \textit{paper} accumulates, and the number of citations that a \textit{scientist} accumulates. To this end, we introduce the following notation: 

\begin{enumerate}
    \item $c^G_5(p_j)$: The number of citations that paper $p_j$ accumulates 5 years after publication, where ``G'' stands for ``Group'';   
    \item $c^I_5(s_i)$: The average number of citations that scientist $s_i$ accumulates from a paper 5 years after its publication, where ``I'' stands for ``Individual''. More formally:
\begin{equation}\label{eq:individualImpact}
c^I_5(s_i)=\frac{\sum\limits_{p_j\in\mathit{Papers}(s_i)}c^G_5(p_j)}{ |\mathit{Papers}(s_i)|}.
\end{equation}
\end{enumerate}

To improve readability, we will often write  $c^G_5$ instead of $c^G_5(p_j)$ whenever the paper is clear from the context. Similarly, we will write $c^I_5$ instead of $c^I_5(s_i)$ when there is no risk of confusion.

Various studies have demonstrated that the average number of citations per paper changes over time \cite{radicchi2008universality,bornmann2008citation,althouse2009differences,sinatra2016quantifying}. To mitigate this temporal effect, we follow the approach proposed by Sinatra et al.~\cite{sinatra2016quantifying}, and consider an alternative, normalized measure of impact, defined as:
$$
\widetilde{c}^G_5\ =\ \frac{c^G_5(p_j)}{\big<c^G_5\big>_{\mathit{year}(p_j)}},
$$

\noindent where $\big<c^G_5\big>_{\mathit{year}(p_j)}$ denotes the average $c_5$ taken over all papers published in the same year as $p_j$. Similarly, when analyzing the impact of a scientist $s_i$, we considered an alternative, normalized version of $c^I_5(s_i)$, defined as follows:
\begin{equation}\label{eq:individualImpactNormalized}
\widetilde{c}^I_5(s_i)=\frac{\sum\limits_{p_j\in\mathit{Papers}(s_i)}\widetilde{c}^G_5(p_j)}{ |\mathit{Papers}(s_i)|}.
\end{equation}

Considering every paper in the entire MAG dataset, we found that $c^G_5$ and $\widetilde{c}^G_5$ are very strongly correlated, with Pearson's $r=0.965$ and $p<0.0001$ (Supplementary Figure~\ref{fig:c5_vs_normalized_c5} depicts $c^G_5$ against $\widetilde{c}^G_5$ for 500,000 papers chosen uniformly at random). Note that there is no need to repeat this analysis for $\widetilde{c}^I_5$ since it is derived from $\widetilde{c}^G_5$; see Equation~\eqref{eq:individualImpactNormalized}. Based on this finding, all subsequent analysis uses the unnormalized versions, i.e., $c^G_5$ and $c^I_5$, since they seem to be more intuitive and interpretable, as argued in \cite{sinatra2016quantifying}.


\newpage
\section{\hspace{-.5cm}.\hspace{.35cm}Coarsened Exact Matching}\label{sec:CEM}
To establish a causal link between ethnic diversity and scientific impact, we use \emph{coarsened exact matching (CEM)} \cite{iacus2012causal}, a technique used to infer causality in observational studies. Specifically, it matches the control and treatment populations with respect to the confounding factors identified, thereby eliminating the effect of these factors on the phenomena under investigation. In our case, when studying a paper's \textit{group} ethnic diversity, we identified the following confounding factors and bins (we experimented with other binning decision and the results were found to be robust to the binning decisions):
\vspace*{-0.3cm}
\begin{itemize}\itemsep-0.5em
\item \textbf{year of publication}: 5 bins, the first of which contains papers published before 1990; the remaining 4 bins reflect 5-year intervals between 1990 and 2010.
\item \textbf{number of authors}: Each bin corresponds to a single number.
\item \textbf{field of study}: 8 bins, one for each of the main fields of science (see Supplementary Note~\ref{SM:GoogleScholar}).
\item \textbf{authors' impact prior to publication}: 
An author's prior impact is measured as the average number of citations that he/she accumulated per year over the period that precedes the year in which the paper was published. The prior impact of all authors is binned into 3 bins, one corresponding to the top 25\%, one the middle 50\%, and one the lowest 25\%.
\item  \textbf{university rankings}:\footnote{ University rankings are based on the 2017 \emph{``Academic Ranking of World Universities''}, also known as the \textit{``Shanghai ranking''}; see {\tt http://www.shanghairanking.com/ARWU2017.html}} Here, we consider two alternatives:\vspace*{-0.5cm}
    \begin{itemize}
    \item 6 bins, corresponding to the rank of the \textbf{highest-ranked} university; this rank falls in one of the following: 1-100; 101-200; 201-300; 301-400; 401-500; $>$500.
    \item 6 bins, corresponding to \textbf{average rank} of all universities in the paper; this average falls in one of the following: 1-100; 101-200; 201-300; 301-400; 401-500; $>$500.
    \end{itemize} 
\end{itemize}
In contrast, when studying an author's \emph{individual} ethnic diversity, we identified these confounding factors and bins (we experimented with other binning decisions and got similar results):
\vspace*{-0.5cm}
\begin{itemize}\itemsep-0.5em
\item \textbf{academic age}: Each bin corresponds to a single academic age.
\item \textbf{number of collaborators}: Each bin corresponds to a single number.
\item \textbf{discipline}: 19 bins, one for each discipline (see Supplementary Note~\ref{SM:TypesofDiversity}).
\item \textbf{university ranking}: 6 bins, corresponding to scientists whose affiliation rank falls in one of the following: 1-100; 101-200; 201-300; 301-400; 401-500; $>$500.
\end{itemize}
Next, we filter the dataset and retain only papers and scientists for which the above confounding factors are known. Throughout the remaining steps of CEM, we will only deal with this filtered dataset. 
We now move on to selecting the treatment set, $T$, and the control sets, $C$. To this end, let $P_i\left(d^G_{\mathit{eth}}\right)$ be the $i^{th}$ percentile of $d^G_{\mathit{eth}}$. Then, when studying \emph{group} ethnic diversity, the treatment and control sets consist of papers for which $d^G_{\mathit{eth}} > P_{100-i}\left(d^G_{\mathit{eth}}\right)$, and $d^G_{\mathit{eth}} \leq P_i\left(d^G_{\mathit{eth}}\right)$, respectively, where $P_i\left(d^G_{\mathit{eth}}\right)$ denotes the $i^{th}$ percentile of $d^G_{\mathit{eth}}$. This process is repeated using $i=10, 20, 30, 40, 50$, corresponding to progressively larger gaps in ethnic diversity between the two populations. Thus, if ethnic diversity does indeed increase scientific impact, we would expect to find a significant difference in impact between the two populations, and expect the difference to increase in tandem with the aforementioned gap in diversity. The same process was carried out for \emph{individual} ethnic diversity, but with $d^I_{\mathit{eth}}$ instead of $d^G_{\mathit{eth}}$. This CEM process is illustrated in Supplementary Figure~\ref{fig:CEM_infographic}, whereas the distributions of the confounding factors in both treatment and control groups (before the CEM process) are depicted in Supplementary Figures \ref{fig:confoundingFactorsDistributions:group} and \ref{fig:confoundingFactorsDistributions:individual}.

The CEM results for \textit{group} ethnic diversity can be found in Table~2 of the main article, whereby the confounding factor ``university ranking'' corresponds to the \emph{average rank} of all universities in the paper. Similarly, the CEM results for \textit{individual} ethnic diversity can be found in Table~3 of the main article, whereby the ``university ranking'' corresponds to the rank of the scientist's affiliation. In contrast, here we present the CEM results for group ethnic diversity whereby ``university ranking'' corresponds to the rank of the \emph{highest-ranked} university in the paper; see Supplementary Table~\ref{tab:CEMGroup}. Similar broad trends can be observed, compared to Tables~2 and 3.


\newpage
\section{\hspace{-.5cm}.\hspace{.35cm}The Relationship between University Rankings and Ethnic Diversity}\label{sec:relationship:ranking:ethnicity}

\noindent In this Supplementary Note, we investigate whether the correlation between ethnic diversity and research impact is due to higher-ranked universities attracting top students with diverse ethnic backgrounds from abroad. University rankings were based on the 2017 ``Academic Ranking of World Universities'', also known as the ``Shanghai ranking''.\footnote{ {\tt http://www.shanghairanking.com/ARWU2017.html}} Supplementary Figure~\ref{fig:unirank_vs_groupdiv}  depicts the following while controlling for the number of authors per paper:

\begin{itemize}
\item Group ethnic diversity against the following university rankings: 1, 2, ..., 99, 100;
\item Group ethnic diversity against the following university ranking bins: 1-100; 101-200; 201-300; 301-400; 401-500;
\end{itemize}

In contrast, Supplementary Figure~\ref{fig:unirank_vs_inddiv} depicts the following while controlling for the number of collaborators per scientist:
\begin{itemize}
\item Individual ethnic diversity against the following university rankings: 1, 2, ..., 99, 100;
\item Individual ethnic diversity against the following university ranking bins: 1-100; 101-200; 201-300; 301-400; 401-500.
\end{itemize}

As anticipated, the correlation between ethnic diversity and university ranking is negative and significant in all cases ($p<0.001$), e.g., a university ranked $10^{\textnormal{th}}$ produces, on average, papers with greater ethnic diversity than another ranked $80^{\textnormal{th}}$. Nevertheless, even when controlling for university ranking, the relationship between ethnic diversity and scientific impact persists, as was shown in Supplementary Note~\ref{sec:CEM}.




\clearpage
{\noindent\Large \textbf{Supplementary References}}
\vspace{-1cm}